\documentclass[12pt,a4paper]{article}

\usepackage[T1]{fontenc}
\usepackage[utf8]{inputenc}
\usepackage{amsmath,amssymb}
\usepackage{graphicx}
\usepackage{xcolor}
\usepackage{hyperref}
\usepackage{natbib}
\usepackage{booktabs}
\usepackage{placeins}
\usepackage{lineno}
\usepackage{geometry}
\usepackage{float}

\newlength{\panelheight}

\title{\textbf{DISTORTED-WAVE CALCULATIONS OF ATOMIC IONIZATION BY BARE IONS AND ANTIPROTONS}}

\author{
Dar\'io M. Mitnik$^{1}$,
Jorge E. Miraglia$^{1}$,
Claudia C. Montanari$^{1}$\\[6pt]
$^{1}$ Instituto de Astronom\'ia y F\'isica del Espacio,\\
CONICET and Universidad de Buenos Aires,\\
1428 Buenos Aires, Argentina\\[4pt]
\textit{Correspondence: dmitnik@df.uba.ar}
}

\date{}
\begin{document}

\maketitle

\begin{abstract}
We present \textit{ab initio} calculations of atomic ionization by bare ions and antiprotons within the continuum distorted-wave eikonal initial-state (CDW-EIS) model. The dataset covers atomic targets from H to Kr, impacted by bare ions over a broad range of projectile charges and by antiprotons. Particular attention is given to elements in periods 3 and 4 of the periodic table, which are relevant to pigment analysis and technologically important materials. The resulting subshell-resolved total ionization cross sections are made openly available through an online repository covering a broad range of projectile--target systems, occupied initial subshells, and projectile velocities. 
%We analyze systematic trends in the cross sections, including their dependence on projectile charge and velocity, binding-energy scaling, and charge-sign effects in proton and antiproton impact. 
To extend the practical coverage of the database, we developed a neural-network surrogate, CDWEIS-NN, trained on the subset of the database with projectile charge  
$|Z_\mathrm{P}|\leq 8$. The network extends the CDW-EIS results to additional projectile--target systems and impact energies, reproducing both held-out calculations and the experimental benchmarks considered in this work with satisfactory accuracy.
\end{abstract}


%\linenumbers
%%%%%%%%%%%%%%%%%%%%%%%%%%%%%%%%%%%%%%%%%%%%%%%%%%%%%%%%%%%%%%%%%%%%%%%%
%%%%%%%%%%%%%%%%%%%%%%%%%%%%%%%%%%%%%%%%%%%%%%%%%%%%%%%%%%%%%%%%%%%%%%%%
\section{Introduction}
\label{sec:introduction}

Particle-induced X-ray emission (PIXE) \cite{Johansson_PIXE} is widely used in art and cultural-heritage studies \cite{MAGALINI2025,RIZZUTTO2014,Sottili2022} and in the analysis of technological materials, including the characterization of advanced steels \cite{Latifi2024,Zhou_2016}. Its nondestructive character and quantitative elemental sensitivity make PIXE particularly valuable in these applications. Accurate inner-shell ionization cross sections, mainly for the $K$ and $L$ shells and, for heavy elements, also for the $M$ shell, are required for reliable PIXE analysis. Such cross sections are obtained from experimental compilations \cite{PAUL1982,PAUL1989,PAUL1993,MIRANDA2014} or from semiempirical models based on the plane-wave Born approximation, notably ECPSSR and its extensions \cite{Brandt-Lapicki,LAPICKI2020}, as implemented, for example, in ISICS \cite{CIPOLLA2011}, whose accuracy varies across collision regimes. Cross sections inferred from characteristic X-ray yields depend on the fluorescence and relaxation data used in the conversion. Nonperturbative approaches, such as close-coupling or classical-trajectory methods \cite{Kirchner_2011}, are required at low velocities but are computationally demanding for multielectron targets.

Atomic targets in periods 3 and 4 of the periodic table, from Na to Kr, include elements of particular practical interest: Al and Si occur in pigments and technological materials, while the transition metals from Sc to Zn are relevant to a variety of analytical and materials-science applications \cite{Sottili2022,Latifi2024,Zhou_2016,Sousa2024}. Their multielectron and multiorbital structure makes a systematic \textit{ab initio} treatment demanding. In this work, we address this problem with the continuum distorted-wave eikonal initial-state (CDW-EIS) model \cite{Crothers1983,Fainstein1988,Rivarola1996,Miraglia2008,Miraglia2009,Montanari_2010}. We calculate subshell-resolved ionization cross sections for atomic targets impacted by bare ions over a broad range of projectile charges, as well as by antiprotons, thereby providing values for an extensive set of projectile--target systems and occupied $nl$ subshells. The present calculations span projectile velocities $v=1$--20~a.u., i.e., energies up to about 10~MeV/u, which include the proton energies typically used in PIXE analysis.

The CDW-EIS approach has been extensively tested against experiment in previous applications, mainly focused on gases and biological molecules \cite{Miraglia2009,Montanari_2010,Agnihotri2012,MONTANARI2017_NIMB,Tilda-V,Tribedi_2022,Quinto2025,Mendez2020b}; its formulation and earlier tests are reviewed in Ref.~\cite{Rivarola1996}. The calculations reported here, however, require a substantial computational effort. A central outcome of the present work is therefore an open-access database \cite{CDWEISdatabase} containing tabulated cross sections for a broad range of projectiles, targets, and initial bound states. 
%We also use the database to examine systematic trends and binding-energy scaling with projectile charge and velocity.

Finally, we develop the CDWEIS-NN neural-network surrogate to learn the systematic behavior of the calculated data and extend them to projectile--target combinations and impact energies for which direct CDW-EIS results are not available. Its performance is assessed first against held-out CDW-EIS calculations and, as a final external test, against available experimental measurements. 

The manuscript is organized as follows. Section~\ref{sec:cdweis} summarizes the theoretical model. 
Section~\ref{sec:database} describes the CDW-EIS database and analyzes systematic trends in the ionization cross sections. 
Section~\ref{sec:expcomparison} compares the CDW-EIS calculations with available experimental data.
Section~\ref{sec:cdwNN} presents the CDWEIS-NN model, its validation, and representative predictions. 
Finally, Section~\ref{sec:conclusions} draws the main conclusions and outlines future perspectives. 
Atomic units (a.u.) are used throughout this work.

%%%%%%%%%%%%%%%%%%%%%%%%%%%%%%%%%%%%%%%%%%%%%%%%%%%%%%%%%%%%%%%%%%%%%%%%
%%%%%%%%%%%%%%%%%%%%%%%%%%%%%%%%%%%%%%%%%%%%%%%%%%%%%%%%%%%%%%%%%%%%%%%%
\section{The CDW-EIS Model}
\label{sec:cdweis}

The CDW-EIS model describes ionization induced by a bare projectile of nuclear charge $Z_\mathrm{P}$ colliding with an atomic target of nuclear charge $Z_\mathrm{T}$. It includes projectile distortion in both the initial and final channels and provides a higher-order description of the collision. The approach has been tested against experimental data in a variety of systems \cite{Agnihotri2012,MONTANARI2017_NIMB,Tilda-V,Tribedi_2022,Quinto2025}.
As a complementary statistical characterization, an earlier set of subshell-resolved CDW-EIS ionization cross sections was found to exhibit agreement with Benford's law \cite{Miraglia2022}.

The total ionization cross section of an electron initially in the $nlm$
state is given by the four-dimensional integral

\begin{equation}
\sigma_{nlm}(v)=\frac{(2\pi )^{2}}{v^{2}}\int d\vec{k} 
\int d\vec{\eta}\ \left\vert T_{nlm}(\vec{k},
\vec{\eta})\right\vert ^{2}  
\label{eq:sigmanlm}
\end{equation}
where $T_{nlm}(\vec{k},\vec{\eta})$ is the transition
matrix, $\vec{\eta}$ is the component of the momentum transfer
perpendicular to the incident velocity $\vec{v}$, and $\vec{k}$ is the momentum of the
emitted electron. For bare ionic projectiles, which are treated as point charges, the integration over $\vec{\eta}$ extends to infinity.

Within the CDW-EIS model, the $T$-matrix element reads \cite%
{Miraglia2008,MONTANARI2017_NIMB} 
\begin{equation}
T_{nlm}^{CDWEIS}(\vec{k},\vec{\eta})= 
-\frac{1}{(2\pi)^{3/2}} \ \vec{W}(\vec{p}) \cdot \vec{G}_{\vec{k},nlm}(\vec{p})
\, ,  
\label{eq:Tcdweis}
\end{equation}
with 
\begin{eqnarray}
\vec{G}_{\vec{k},nlm}(\vec{p}) &=& 
\int d\vec{r} \; \left[ \vec{\nabla}\varphi_{\vec{k}}^{-}(\vec{r})\right] ^{\ast}\, e^{i\;\vec{p} \cdot \vec{r}} \ \varphi_{nlm}(\vec{r}) 
\, ,  
\nonumber \\
\vec{W}(\vec{p}) &=& 
N_{\xi_{p}} \, \int \frac{d\vec{r}}{(2\pi )^{3/2}} \ 
E_{\vec{v}}^{+}(\vec{r}) \ e^{-i\;\vec{p}\cdot \vec{r}} \ 
\vec{\nabla}F_{\vec{k}_{p}}^{-^{\ast}}(\vec{r}) \, ,  
\nonumber \\
E_{\vec{v}}^{+}(\vec{r}) &=& 
\exp \left[ -i\xi _{v} \ln (v r+\vec{v} \cdot \vec{r})\right] 
\, , 
\nonumber \\
F_{\vec{k}_{p}}^{-}(\vec{r}) &=& 
_{1}F_{1}(-i\xi_{p},1,-ik_{p} \ r-i\vec{k}_{p} \cdot \vec{r})
\, , 
\nonumber \\
N_{\xi_{p}} &=& 
\exp (\xi _{p}/2) \ \Gamma (1-i\xi _{p}) 
\, , 
\end{eqnarray} 
where $\xi _{v}=Z_{P}/v$, $\xi _{p}=Z_{P}/k_{p},\vec{\ k}_{p}=%
\vec{k}-\vec{v}$, $_{1}F_{1}$ is the confluent
hypergeometric function, and $\vec{p}$ is the projectile momentum
transferred, which can also be expressed as $\vec{p}=(%
\vec{\eta },p_{m})$, with $p_{m}$ being the minimum momentum
transfer along the direction of $\vec{v}$, $p_{m}=(\varepsilon
_{f}-\varepsilon _{i})/v$, and $\varepsilon _{i}$ ($\varepsilon _{f}$) the
initial (final) electron energy.

Both bound, $\varphi_{nlm}$, and continuum states, $\varphi_{\vec{k}}^{-}$, are numerical solutions of the same Hamiltonian. They are obtained with the \textsc{radialf} code developed by Salvat and collaborators \cite{SALVAT1995,SALVAT2019}. 
The central potentials required by \textsc{radialf} are derived from the Roothaan--Hartree--Fock wave functions, expanded in Slater-type orbitals, tabulated by Bunge \textit{et al.}~\cite{BUNGE1993}, using the Depurated Inversion
Method (DIM) \cite{Mendez2018_DIM}. By construction, this potential ensures
that the bound and continuum wave functions are properly orthogonal.

We performed the radial integration numerically and integrated the angular
variables using the vector-coupling technique \cite{Edmonds1985}. The radial
grid consisted of several thousand points, depending on the spatial extent
of the initial $nlm$-state. 
The continuum state $\varphi_{\vec{k}}^{-}$ was expanded in spherical harmonics with orbital angular momenta up to $l=30$.

%%%%%%%%%%%%%%%%%%%%%%%%%%%%%%%%%%%%%%%%%%%%%%%%%%%%%%%%%%%%%%%%%%%%%%%%
%%%%%%%%%%%%%%%%%%%%%%%%%%%%%%%%%%%%%%%%%%%%%%%%%%%%%%%%%%%%%%%%%%%%%%%%
\section{The CDW-EIS Database}
\label{sec:database}

\begin{figure}[H]
    \centering
    \includegraphics[width=0.90\textwidth]{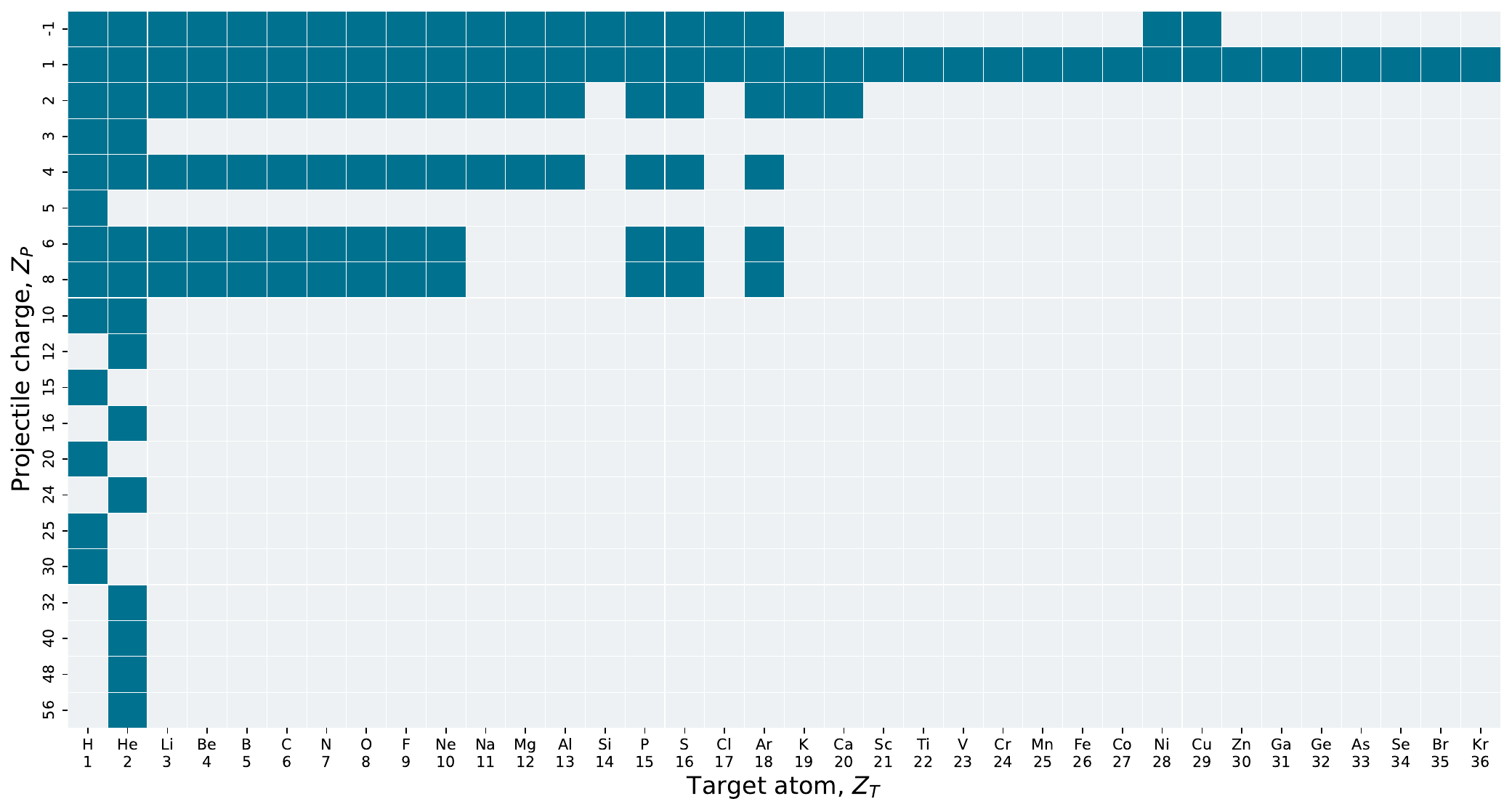}
    \caption{Projectile--target pairs currently available in the CDW-EIS database \cite{CDWEISdatabase}. The vertical coordinate indicates the projectile charge, $Z_\mathrm{P}$, and the horizontal coordinate denotes the target atomic number, $Z_\mathrm{T}$.}
    \label{fig:database-coverage}
\end{figure}

The present CDW-EIS database comprises 132 projectile--target systems. The currently available systems are displayed in Fig.~\ref{fig:database-coverage}. For each projectile--target pair, ionization cross sections were calculated for all occupied $nl$ subshells at a set of projectile velocities in the range $v=1$--$20$ a.u., typically around twelve points per system. The upper limit of this range differs among systems, since the calculations were carried out for different purposes and their convergence at high velocities requires a large number of partial waves.
The present release contains 6239 tabulated subshell-resolved cross sections.

All calculated cross sections are openly available through the CDW-EIS database \cite{CDWEISdatabase}. The interactive Plotly plots \cite{plotly2015} allow users to select curves, zoom, inspect individual points, and export selected data as CSV. The version number and release date of the database should be quoted when the data are used.
Each entry specifies the projectile and target atomic numbers, $Z_\mathrm{P}$ and $Z_\mathrm{T}$, the subshell quantum numbers $n$ and $l$, the subshell occupation number $N_{nl}$, the binding energy $E_{nl}$, the projectile velocity $v$, and the ionization cross section $\sigma_{nl}$. The tabulated quantity $\sigma_{nl}$ includes the occupation number of the corresponding subshell.

The total subshell ionization cross sections were obtained by integrating the differential cross sections over the energy and emission angle of the ejected electron. The repository is intended as a growing resource and will be extended as additional projectile--target combinations and collision energies become available. The corresponding differential cross sections are not currently included in the online database but are available from the corresponding author upon request, and may be incorporated in future releases.

%%%%%%%%%%%%%%%%%%%%%%%%%%%%%%%%%%%%%%%%%%%%%%%%%%%%%%%%%%%%%%%%%%%%%%%%
\subsection{Applicability and limitations}
\label{subsec:applic}

The calculations are based on a nonrelativistic description of the target electronic structure. In particular, the bound and continuum wave functions and the binding energies are obtained from the Schr\"odinger equation rather than the Dirac equation. Relativistic effects in the target become increasingly important at large $Z_\mathrm{T}$, especially for inner-shell electrons, and are not included in the present calculations. For the heaviest targets of the database (Fe, Cu, and Kr), the corrections to the $1s$ binding energies are of the order of $(\alpha Z_\mathrm{T})^2/4 \approx 1$--2\%, consistent with Hartree--Fock and Dirac--Hartree--Fock orbital energies \cite{pyscf,Sun2020PySCF} and with our previous calculations for Zr and Nb \cite{mendez2019relativistic}, in which the wave functions were not substantially modified. Previous CDW-EIS calculations for rare gases up to Xe, involving inner-shell multiple ionization, also gave very good agreement with experiment \cite{Montanari_2012}. We therefore consider the nonrelativistic description adequate for the present targets ($Z_\mathrm{T}\leq 36$).

The projectile charge and velocity also affect the applicability of the collision model. The Presniakov scaling rule \cite{Janev:80,Miraglia:21}, originally discussed for hydrogen, provides a useful guide to the relevant collision regime. For an electron with binding energy $I_{nl}$, the approximate condition can be written as
\begin{equation}
\frac{v^2}{|Z_P| \, \sqrt{2 \, I_{nl}}  } > 1
\label{eq:scaling}
\end{equation}
where $I_{nl}$ is the positive binding energy
\begin{equation}
I_{nl}=|E_{nl}| \, .
\label{eq:Inl}
\end{equation}
This relation should not be regarded as a strict boundary for the CDW-EIS approximation, but rather as an indication of the combinations of projectile charge, velocity, and target binding energy for which the results are expected to be more reliable. In particular, calculations at low velocities and for highly charged projectiles require greater caution.

These considerations guided the choice of collision systems shown in Fig.~\ref{fig:database-coverage}. Most of the database covers antiprotons and bare ions with $1 \leq Z_\mathrm{P} \leq 8$, targets with $1 \leq Z_\mathrm{T} \leq 19$, and incident velocities in the range $1 \leq v \leq 20$ a.u. Additional calculations extend to $Z_\mathrm{P}=56$ for He and to $Z_\mathrm{T}=36$ for selected projectiles. These results broaden the coverage of the repository, although their accuracy may be more limited in some collision regimes.

A small number of calculated cross sections (approximately forty), all at low velocities, exhibit irregularities with respect to the otherwise smooth velocity dependence. Among them, the most noticeable are some $2s$ cross sections, which depart markedly from the systematic behavior of the neighboring points. These points are retained in the online repository but are not included in the training set of the neural-network model described in Section~\ref{sec:cdwNN}. Large deviations between individual calculations and the network predictions can be used to identify entries that may warrant further numerical checks.

%%%%%%%%%%%%%%%%%%%%%%%%%%%%%%%%%%%%%%%%%%%%%%%%%%%%%%%%%%%%%%%%%%%%%%%%
\subsection{Representative ionization cross sections}

\begin{figure}[H]
    \setlength{\panelheight}{0.299\textwidth}
    \begin{minipage}[c]{0.43\textwidth}
        \includegraphics[width=\linewidth]{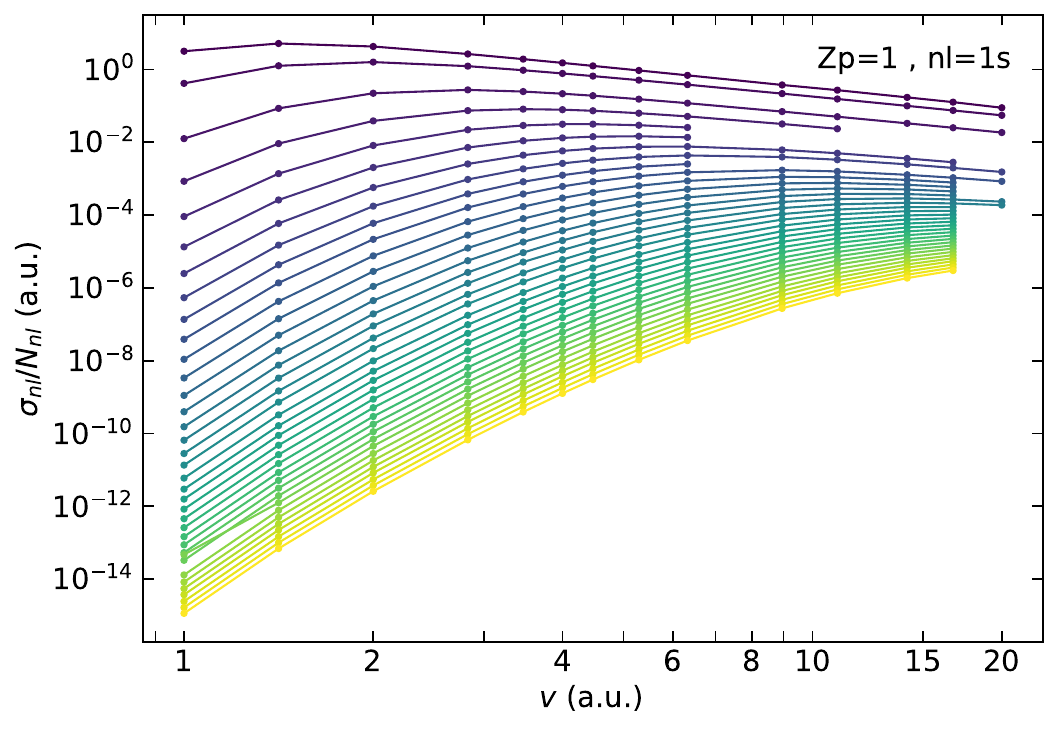}
    \end{minipage}
    \hfill
    \begin{minipage}[c]{0.43\textwidth}
        \includegraphics[width=\linewidth]{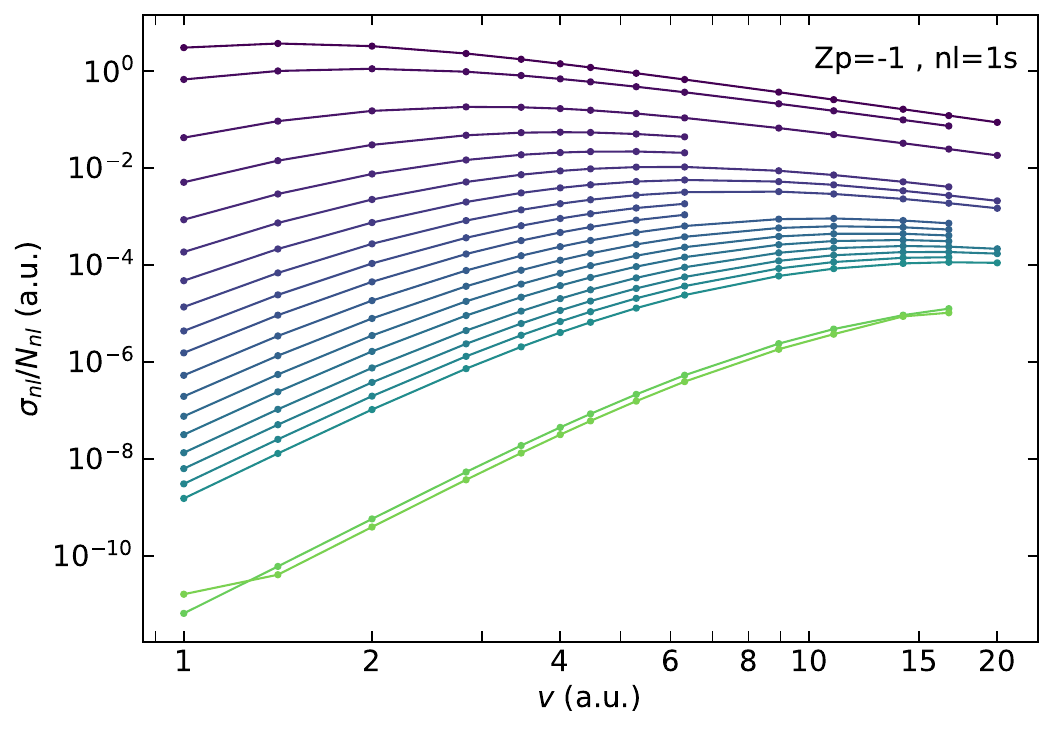}
    \end{minipage}
    \hfill
    \begin{minipage}[c]{0.045\textwidth}
        \includegraphics[height=\panelheight]{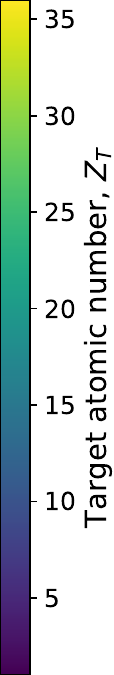}
    \end{minipage}
    \caption{K-shell ionization cross sections per electron as functions of the projectile velocity for different target atoms. Left: proton impact. Right: antiproton impact. The color scale identifies the target atomic number, $Z_\mathrm{T}$.}
    \label{fig:K-charge-sign}
\end{figure}

\begin{figure}[H]
    \setlength{\panelheight}{0.299\textwidth}
    \begin{minipage}[c]{0.43\textwidth}
        \includegraphics[width=\linewidth]{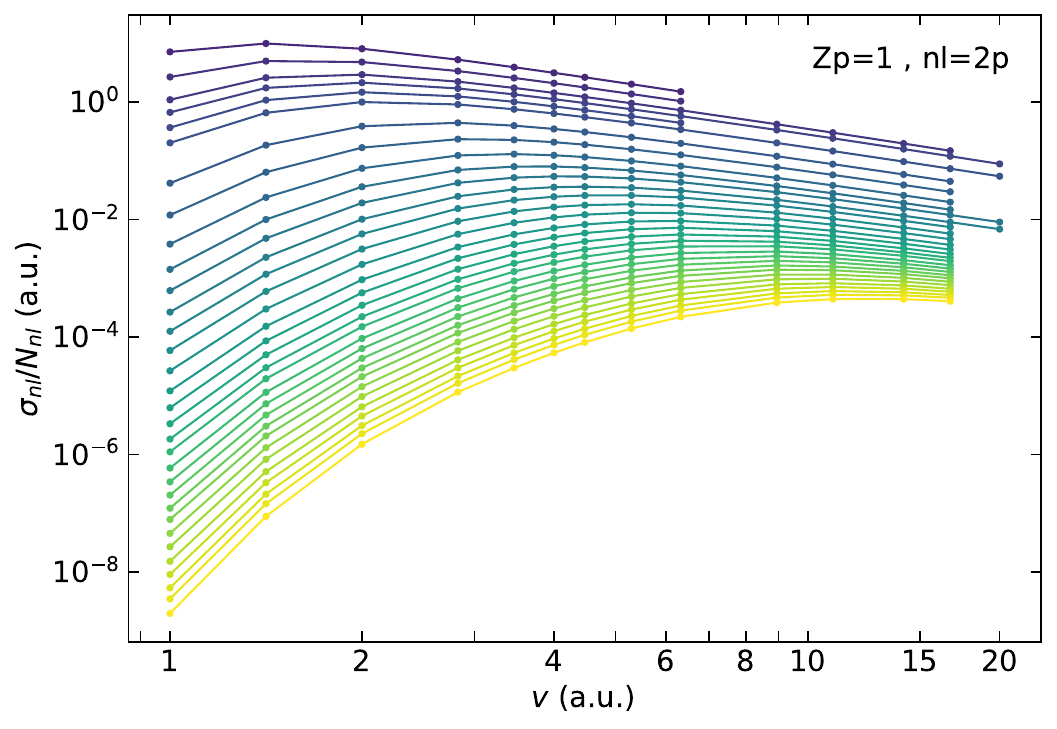}
    \end{minipage}
    \hfill
    \begin{minipage}[c]{0.43\textwidth}
        \includegraphics[width=\linewidth]{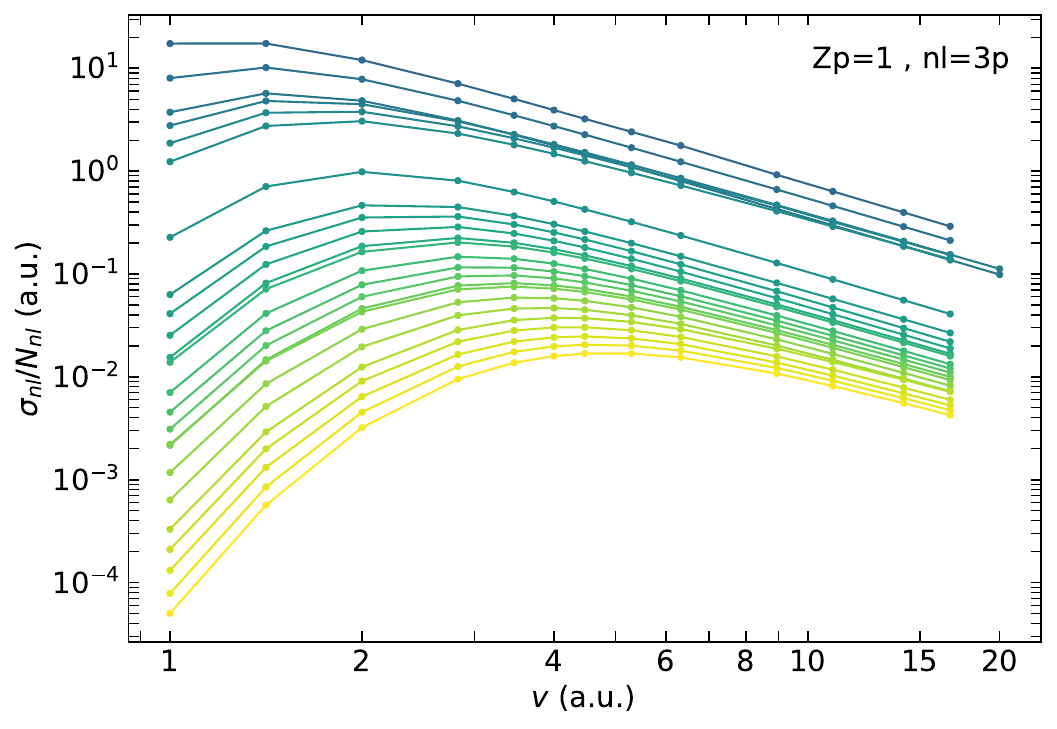}
    \end{minipage}
    \hfill
    \begin{minipage}[c]{0.045\textwidth}
        \includegraphics[height=\panelheight]{figures/target_Z_colorbar.pdf}
    \end{minipage}
    \caption{Proton-induced $2p$ (left) and $3p$ (right) ionization cross sections per electron as functions of the projectile velocity for different target atoms. The color scale identifies the target atomic number, $Z_\mathrm{T}$.}
    \label{fig:proton-outer-shells}
\end{figure}

\begin{figure}[H]
    \setlength{\panelheight}{0.299\textwidth}
    \begin{minipage}[c]{0.43\textwidth}
        \includegraphics[width=\linewidth]{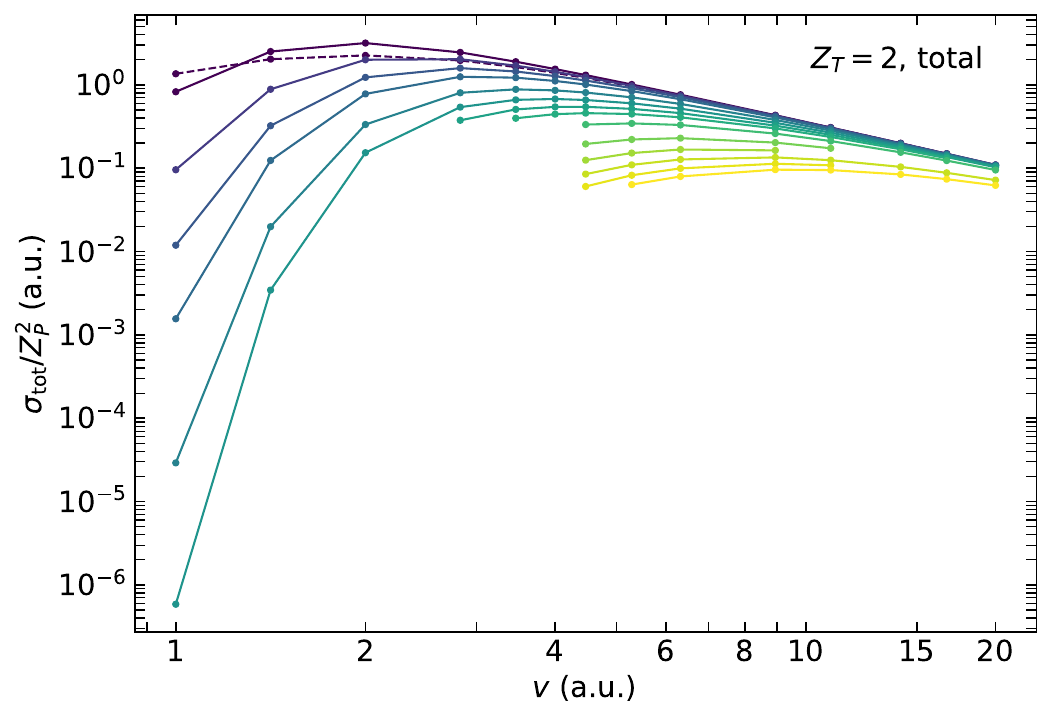}
    \end{minipage}
    \hfill
    \begin{minipage}[c]{0.43\textwidth}
        \includegraphics[width=\linewidth]{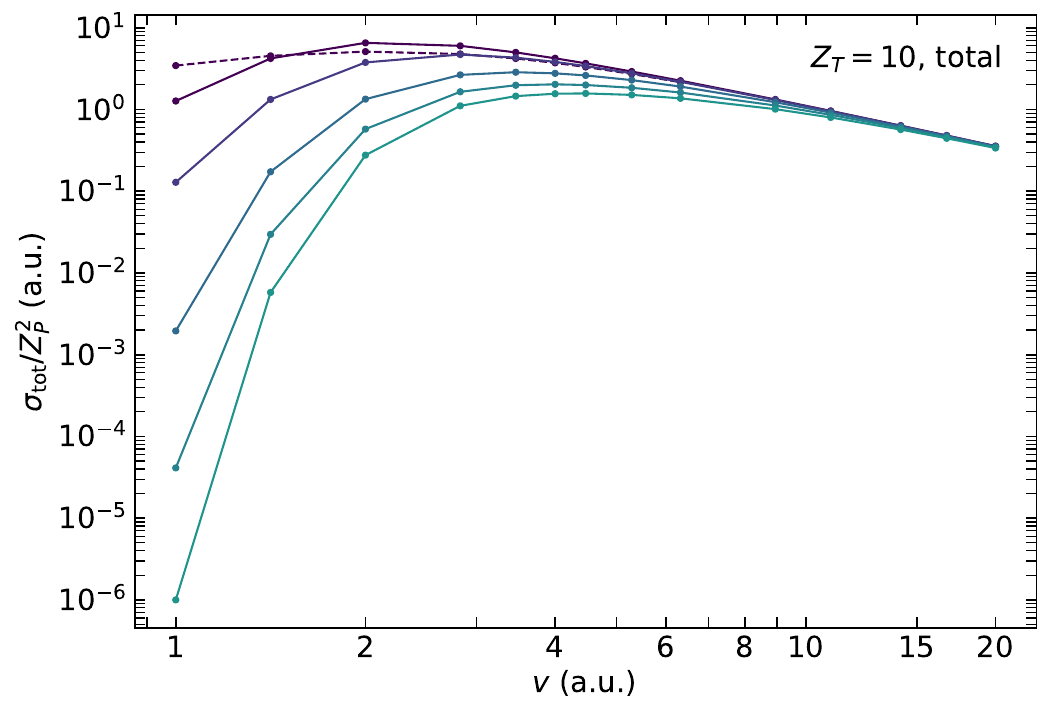}
    \end{minipage}
    \hfill
    \begin{minipage}[c]{0.045\textwidth}
        \includegraphics[height=\panelheight]{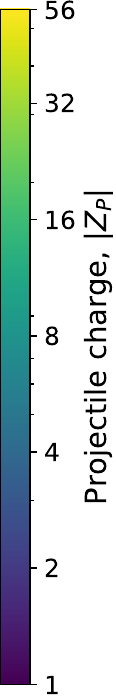}
    \end{minipage}
    \caption{Total ionization cross sections divided by $Z_\mathrm{P}^2$ as functions of the projectile velocity for He (left) and Ne (right). Solid lines correspond to positively charged projectiles and dashed lines to antiprotons. The color scale identifies the magnitude of the projectile charge, $|Z_\mathrm{P}|$.}
    \label{fig:HeandNe}
\end{figure}

\begin{figure}[H]
    \centering
    \includegraphics[width=0.46\textwidth]{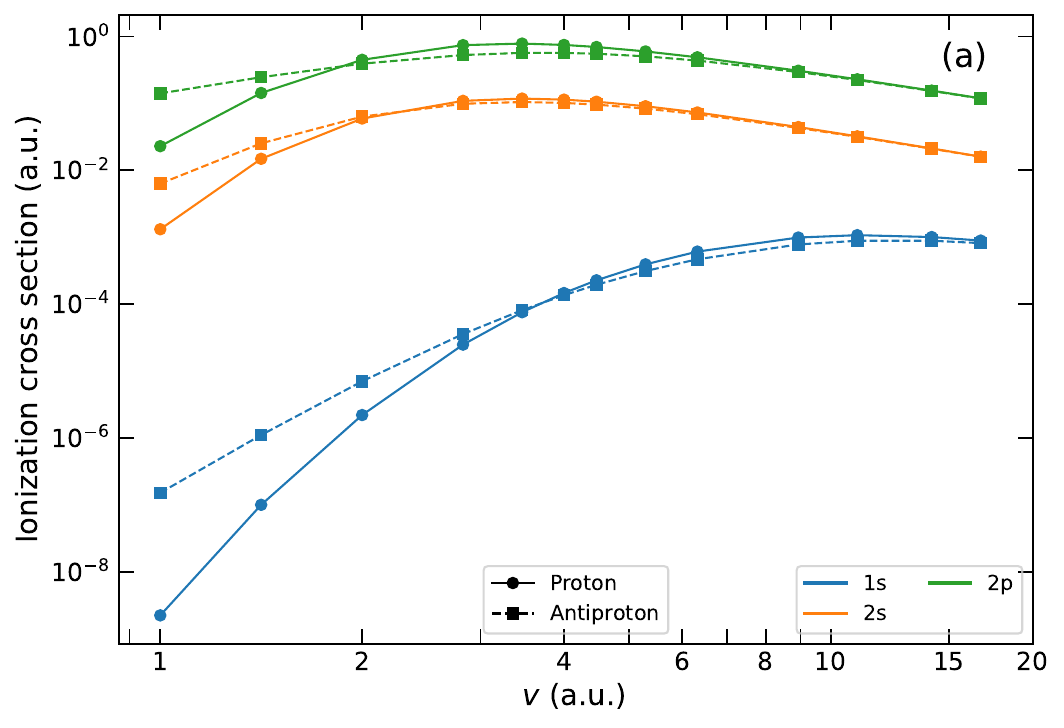}
    \includegraphics[width=0.46\textwidth]{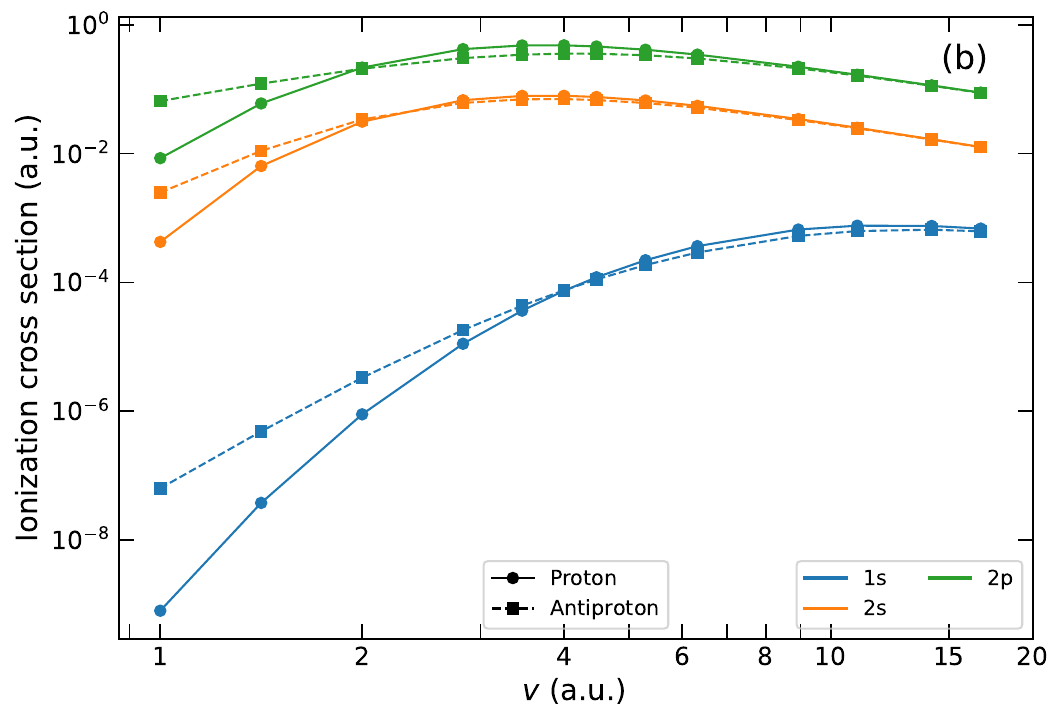} \\
    \includegraphics[width=0.46\textwidth]{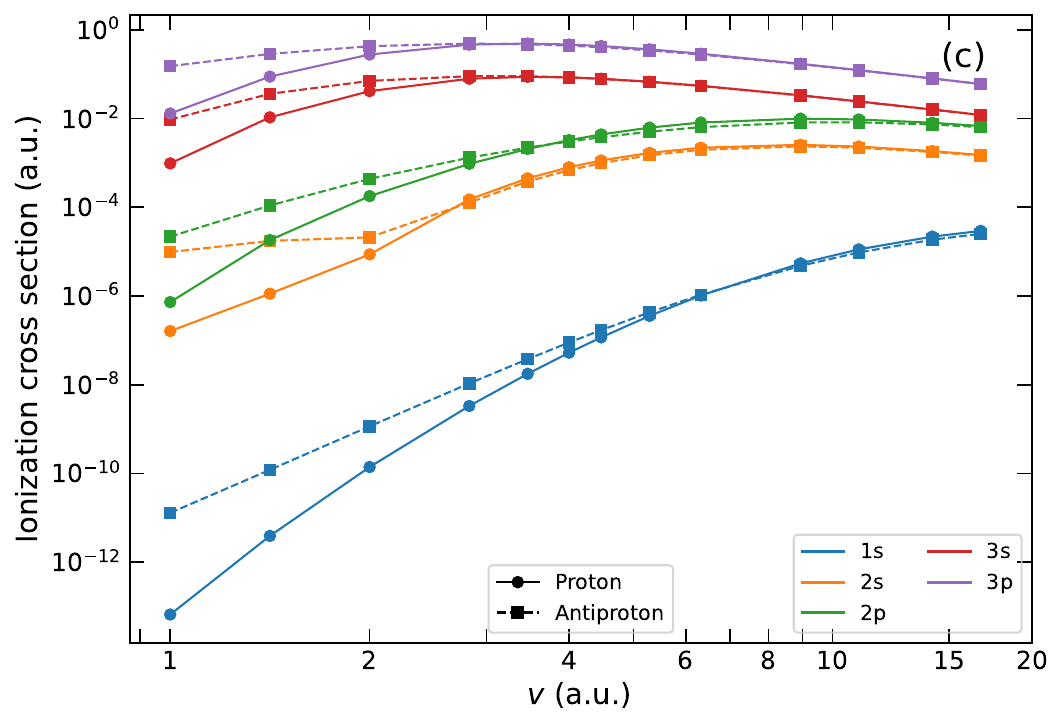}
    \includegraphics[width=0.46\textwidth]{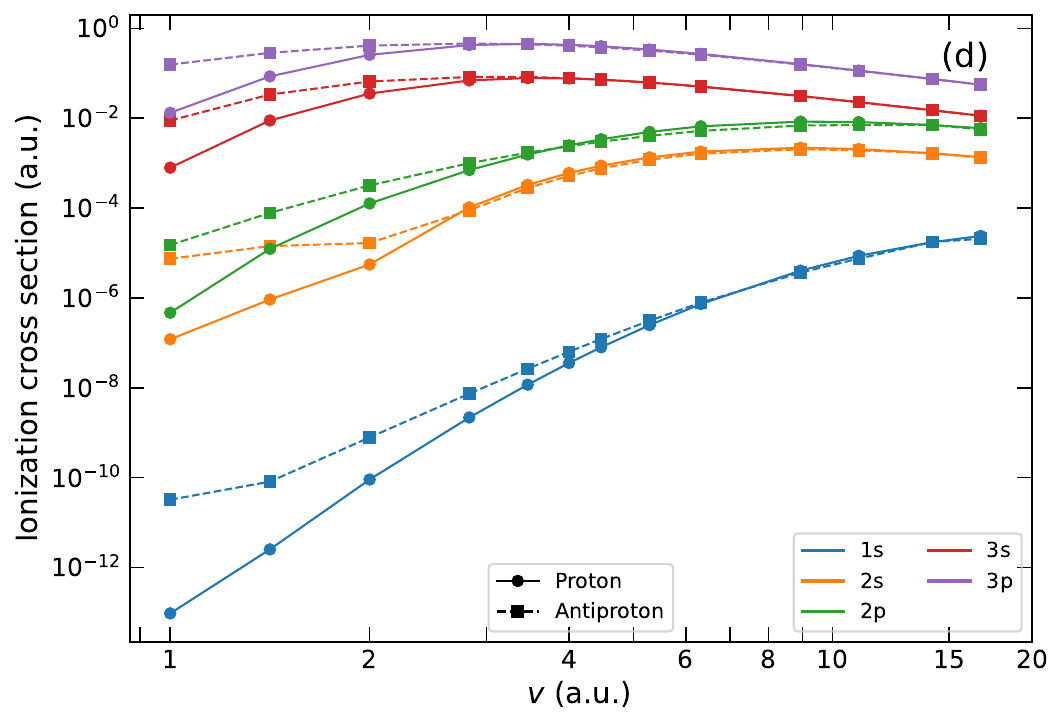}
    \caption{Calculated subshell-resolved ionization cross sections for proton and antiproton impact on (a) Al, (b) Si, (c) Ni, and (d) Cu. The displayed subshells extend from $1s$ through $2p$ for Al and Si, and from $1s$ through $3p$ for Ni and Cu. Solid lines and circles denote proton impact; dashed lines and squares denote antiproton impact.}
    \label{fig:AP-all}
\end{figure}

Figures~\ref{fig:K-charge-sign}--\ref{fig:AP-all} present representative ionization cross sections for different projectile charges, target atoms, and electronic subshells. These examples illustrate the range of systems and the level of detail available in the database. Comparisons with experimental measurements are presented in the following section.

The K-shell cross sections in Fig.~\ref{fig:K-charge-sign} span many orders of magnitude across the target atoms. For the lighter targets, the cross sections reach a maximum within the velocity interval considered and then decrease, whereas for heavier targets they continue to increase over much of this range. Proton and antiproton impact show broadly similar trends, with differences most apparent at low velocities.

Figure~\ref{fig:proton-outer-shells} shows the corresponding results for the $2p$ and $3p$ subshells. Compared with K-shell ionization, these cross sections generally attain larger values and reach their maxima at lower velocities. These differences are mainly associated with the lower binding energies of the outer electrons.

To illustrate the coverage in projectile charge, Fig.~\ref{fig:HeandNe} presents the total ionization cross sections for He and Ne. These are obtained by summing the contributions from all occupied subshells. The cross sections are divided by $Z_\mathrm{P}^2$ to reduce the differences in magnitude between curves and facilitate their visualization; this normalization is not used here to establish a scaling relation. The normalized curves become closer at higher velocities, particularly for Ne, while differences remain for the more highly charged projectiles on He. For the most highly charged projectiles, no calculations were performed at the lowest velocities, where the Presniakov condition of Eq.~\eqref{eq:scaling} is strongly violated; the missing segments in Fig.~\ref{fig:HeandNe} do not indicate that ionization does not take place.

Finally, Fig.~\ref{fig:AP-all} shows the individual subshell contributions for four selected targets, Al, Si, Ni, and Cu. The outer subshells provide the largest contributions in the velocity range shown, while inner-shell cross sections generally rise more strongly with velocity. The comparison also illustrates the differences between proton and antiproton impact, particularly at the lower velocities. These examples emphasize the subshell resolution of the tabulated data, which is retained even when the total ionization cross section is considered.

%%%%%%%%%%%%%%%%%%%%%%%%%%%%%%%%%%%%%%%%%%%%%%%%%%%%%%%%%%%%%%%%%%%%%%%%
%%%%%%%%%%%%%%%%%%%%%%%%%%%%%%%%%%%%%%%%%%%%%%%%%%%%%%%%%%%%%%%%%%%%%%%%
\section{Experimental Validation of the CDW-EIS Calculations}
\label{sec:expcomparison}

The CDW-EIS model has been extensively compared with experiment in previous studies, some of which include systems considered here. The purpose of this section is therefore not to test the model itself, but to verify the present database calculations, performed with a systematic set of target potentials and wave functions. For the experimental cross sections inferred from characteristic X-ray measurements, the fluorescence yields used in the conversion, other vacancy-production channels, and, for solid targets, solid-state effects contribute systematically to the uncertainties.

The reliability of the present CDW-EIS database is assessed by comparison with available measurements of total and subshell-resolved ionization cross sections. The selected benchmarks cover proton, antiproton, and bare-ion impact over a broad range of target atomic numbers and projectile velocities.
For completeness, the figures also show the CDWEIS-NN predictions and their ensemble-spread bands. The construction and validation of the neural-network surrogate are described in Section~\ref{sec:cdwNN}; the emphasis here is on the agreement between the \textit{ab initio} CDW-EIS calculations and experiment.

%%%%%%%%%%%%%%%%%%%%%%%%%%%%%%%%%%%%%%%%%%%%%%%%%%%%%%%%%%%%%%%%%%%%%%%%
\subsection{Proton and antiproton impact on H and He}
\label{subsec:protantiH-He}

Proton--antiproton ionization provides a demanding test for theoretical models because the sign of the projectile charge affects the collision dynamics, particularly at low and intermediate velocities. The CDW-EIS approach has previously shown good agreement with total-ionization measurements for H and He \cite{Rivarola1996}, as well as with multiple-ionization measurements for rare gases \cite{Montanari_2012}.

\begin{figure}[H]
    \centering
    \begin{minipage}{0.47\textwidth}
        \centering
        \includegraphics[width=\linewidth]{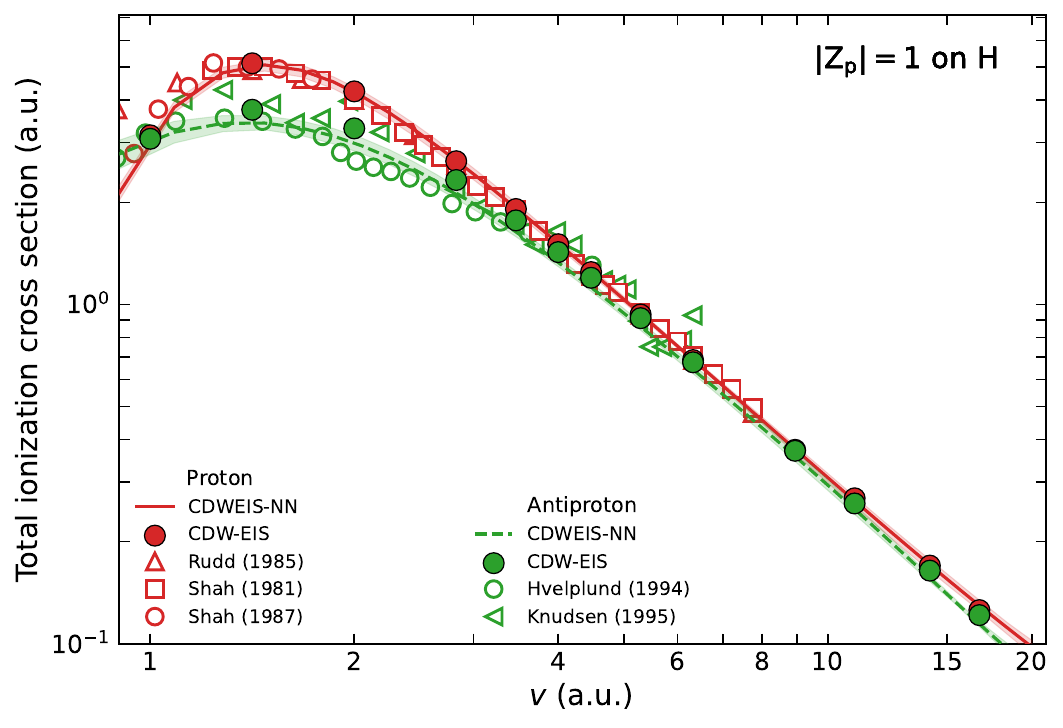}
    \end{minipage}
    \hfill
    \begin{minipage}{0.47\textwidth}
        \centering
        \includegraphics[width=\linewidth]{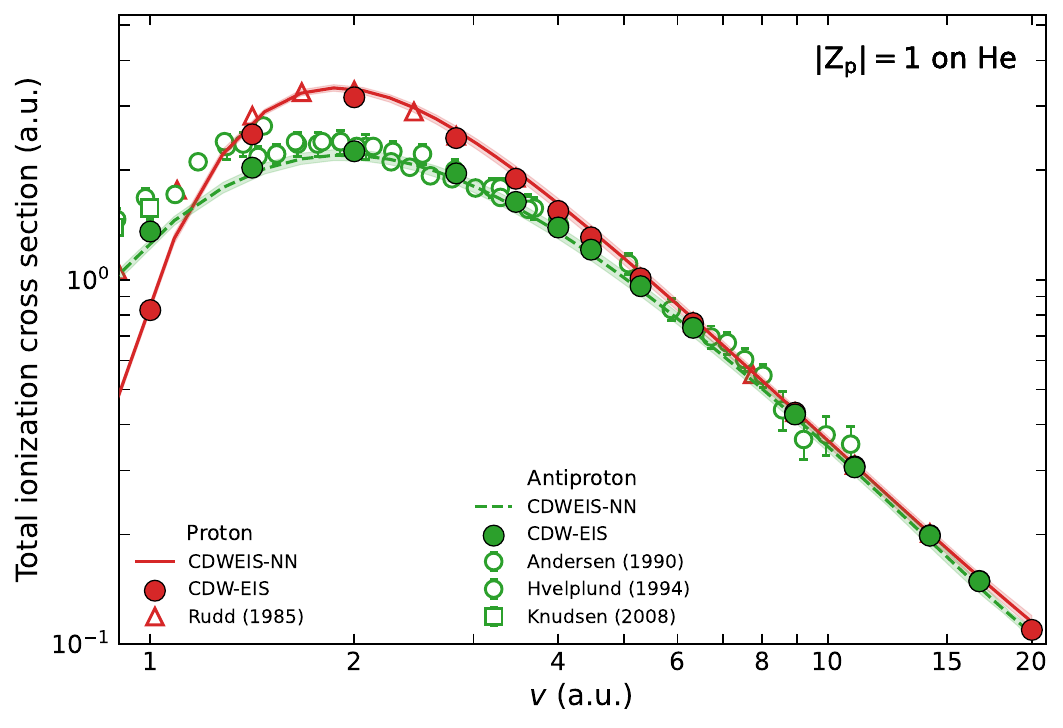}
    \end{minipage}
    \caption{Total ionization cross sections for proton and antiproton impact on hydrogen (left) and helium (right). Filled symbols denote the present CDW-EIS calculations; lines and shaded bands denote the CDWEIS-NN mean predictions and the 16th--84th percentile range of the ensemble members, respectively; the bands should not be interpreted as confidence intervals (see Section~\ref{subsec:Archit}). Antiproton measurements are from \cite{Knudsen1994} for molecular hydrogen, \cite{Knudsen1995} for atomic deuterium, and \cite{Knudsen2008,Knudsen1994,Andersen1990} for helium. Proton-impact measurements are from \cite{Shah1987,Shah1981} for H and \cite{Rudd-RMP1985} for He.}
\label{fig:AP-H-He}
\end{figure}

Figure~\ref{fig:AP-H-He} compares the present calculations with measurements for hydrogen and helium targets. The calculations reproduce the experimentally observed charge-sign dependence, including the crossing of the proton and antiproton cross sections in the low-velocity region. The hydrogen data include measurements for molecular hydrogen and atomic deuterium; these systems are identified separately in the figure and should therefore not be interpreted as identical target species.

For both targets, the CDW-EIS points reproduce the magnitude of the measured cross sections, the location and height of the broad maximum, and the high-velocity decrease. More importantly, the calculation follows the different low-velocity behavior of positively and negatively charged projectiles. This simultaneous description of the two charge signs is a nontrivial test, since it requires the model to reproduce not only the overall ionization scale but also the charge-sign-dependent distortion of the collision dynamics. The close correspondence of the CDWEIS-NN curves with the discrete CDW-EIS points is shown here for reference; its quantitative validation is discussed separately in Section~\ref{sec:cdwNN}.

%%%%%%%%%%%%%%%%%%%%%%%%%%%%%%%%%%%%%%%%%%%%%%%%%%%%%%%%%%%%%%%%%%%%%%%%
\subsection{K-shell ionization}
\label{subsec:Kshell}

\begin{figure}[H]
    \centering
    \begin{minipage}{0.47\textwidth}
        \centering
        \includegraphics[width=\linewidth]{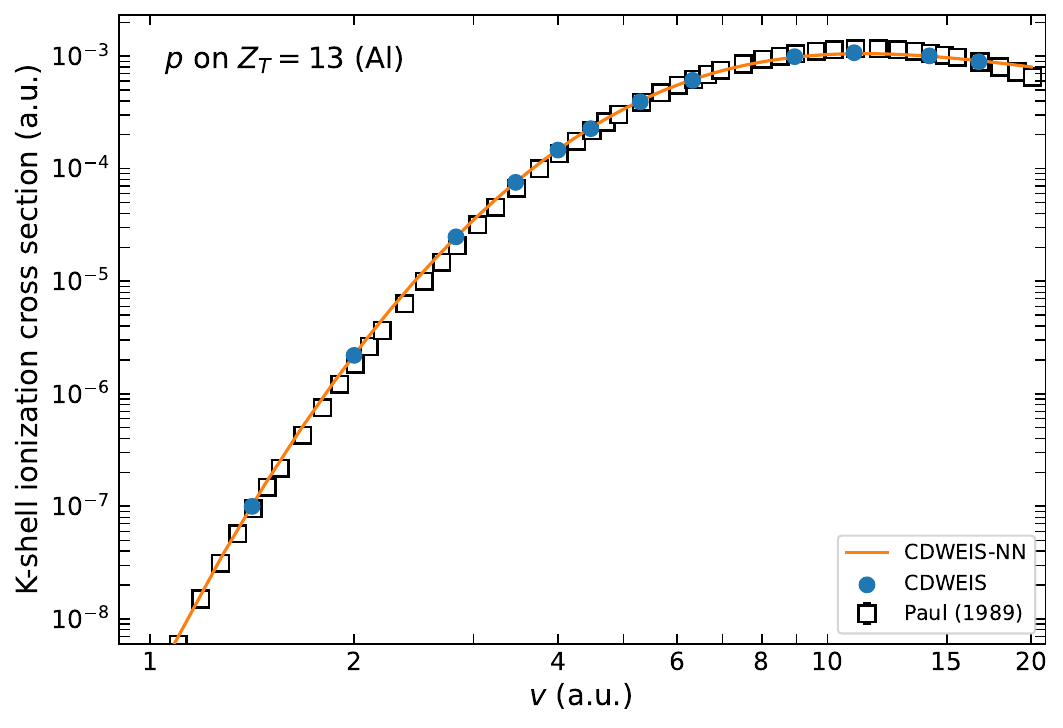}
    \end{minipage}
    \hfill
    \begin{minipage}{0.47\textwidth}
        \centering
        \includegraphics[width=\linewidth]{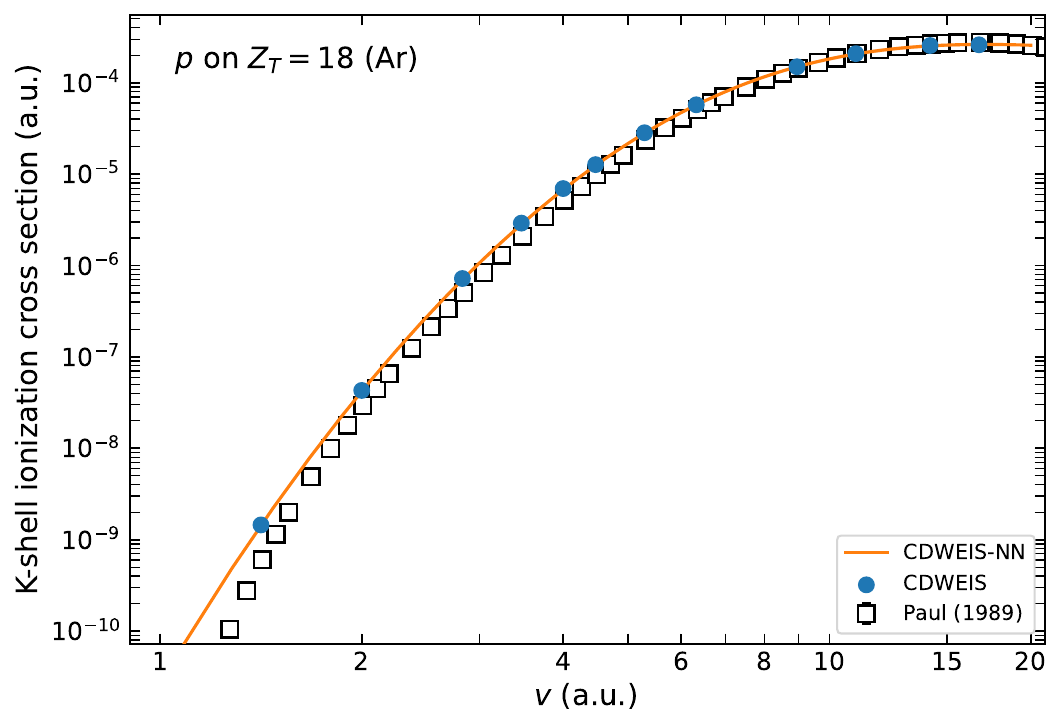}
    \end{minipage}
\caption{K-shell ionization cross sections for proton impact on Al (left) and Ar (right). Filled symbols denote the CDW-EIS calculations; lines and shaded bands denote the CDWEIS-NN mean predictions and the 16th--84th percentile range of the ensemble members, respectively; the bands should not be interpreted as confidence intervals (see Section~\ref{subsec:Archit}); and open symbols denote the experimental data compiled by Paul and Sacher \cite{PAUL1982,PAUL1989}.}
\label{fig:P-Kshell}
\end{figure}

Figure~\ref{fig:P-Kshell} compares the calculated proton-impact K-shell cross sections for Al and Ar with the experimental data compiled by Paul and Sacher \cite{PAUL1982,PAUL1989}. The figure also includes the CDWEIS-NN predictions.

The agreement is excellent over the measured velocity intervals for both Al and Ar. In particular, the direct CDW-EIS calculations reproduce the absolute scale and the velocity dependence of the K-shell cross sections without an empirical normalization. This is a significant benchmark because K-shell ionization probes a deeply bound orbital and is therefore sensitive to the collision dynamics over a substantially different energy scale from that of total ionization in H and He.

\begin{figure}[H]
    \centering
    \begin{minipage}{0.47\textwidth}
        \centering
        \includegraphics[width=\linewidth]{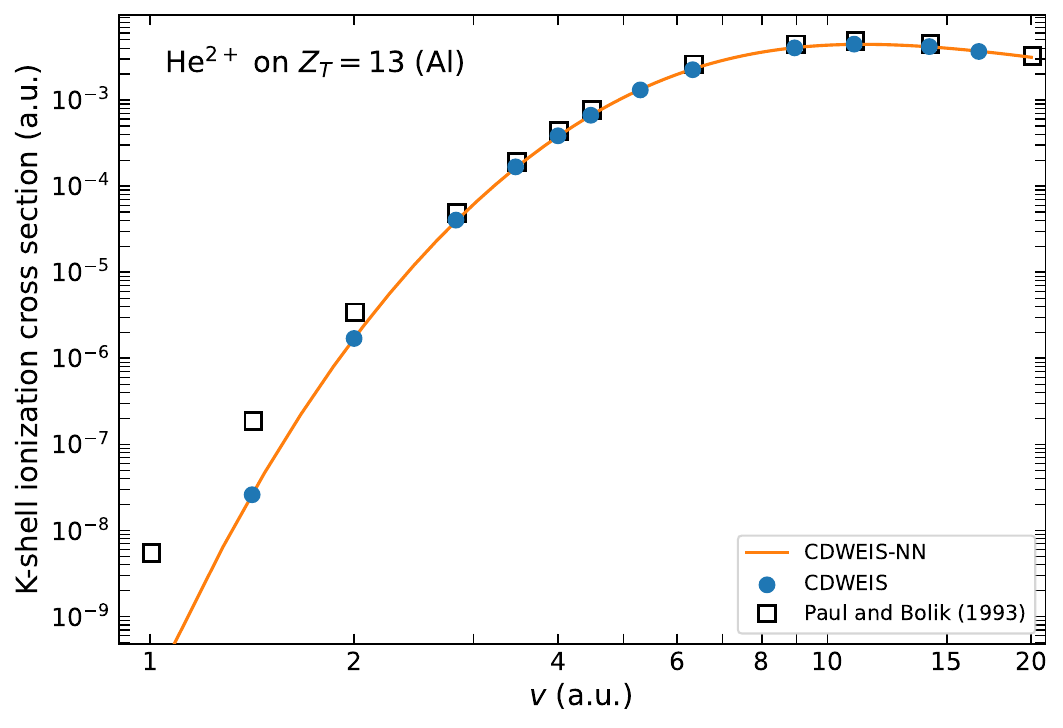}
    \end{minipage}
    \hfill
    \begin{minipage}{0.47\textwidth}
        \centering
        \includegraphics[width=\linewidth]{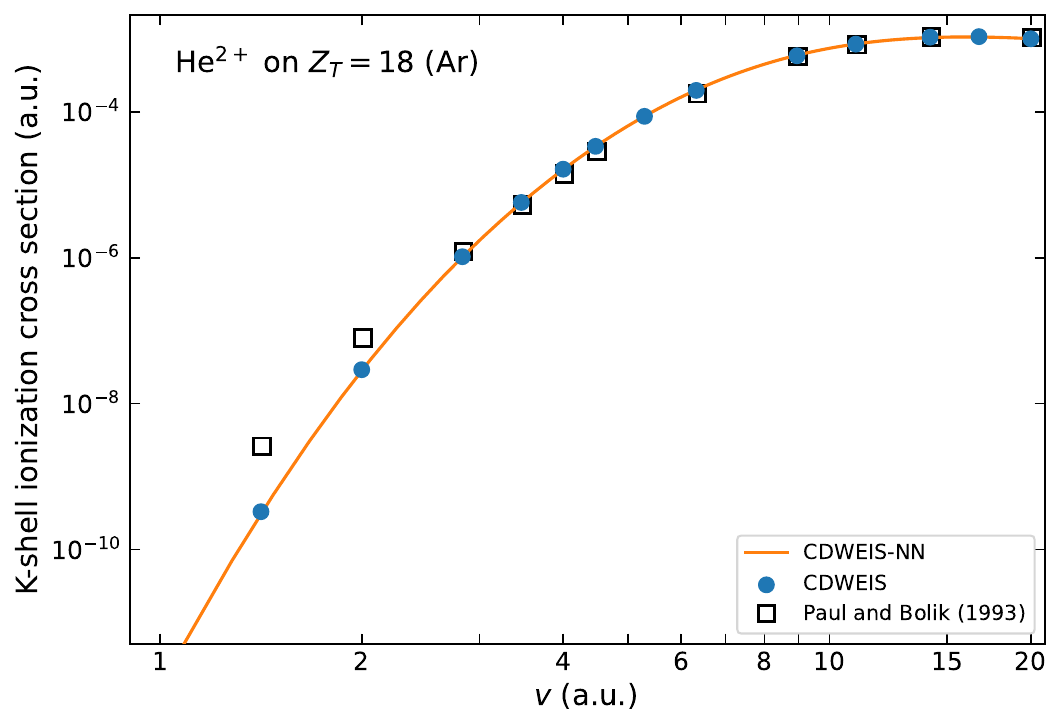}
    \end{minipage}
\caption{K-shell ionization cross sections for He$^{2+}$ impact. Filled symbols denote the CDW-EIS calculations; lines and shaded bands denote the CDWEIS-NN mean predictions and the 16th--84th percentile range of the ensemble members, respectively; the bands should not be interpreted as confidence intervals. Open symbols denote the experimental data compiled by Paul and Bolik \cite{PAUL1993}.}
\label{fig:He2-Kshell}
\end{figure}

Figure~\ref{fig:He2-Kshell} shows K-shell ionization by He$^{2+}$ impact, compared with the measurements compiled by Paul and Bolik \cite{PAUL1993}.

The quality of the agreement is maintained for the doubly charged projectile. Thus, the comparison tests the calculation not only for a different target range but also for a larger projectile perturbation. The CDW-EIS results follow the experimental values for both targets across the measured region, supporting the use of the database for K-shell ionization by bare ions with different charges.

%%%%%%%%%%%%%%%%%%%%%%%%%%%%%%%%%%%%%%%%%%%%%%%%%%%%%%%%%%%%%%%%%%%%%%%%
\subsection{L-shell ionization}

\begin{figure}[H]
\centering
\includegraphics[width=0.47\textwidth]{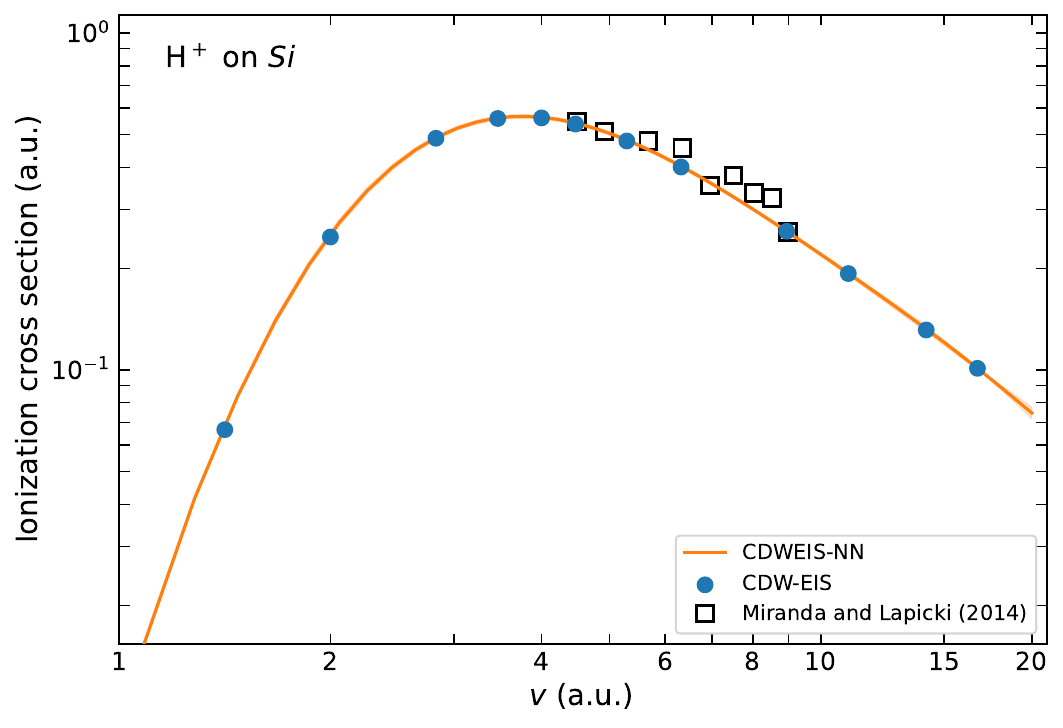}
\includegraphics[width=0.47\textwidth]{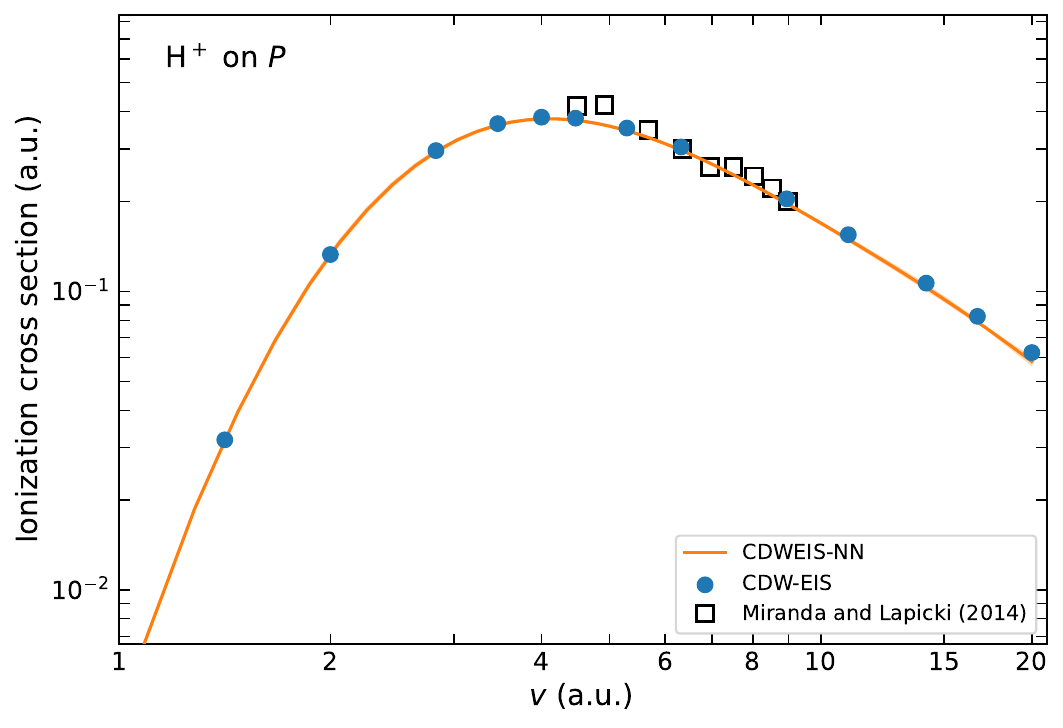}\\
\includegraphics[width=0.47\textwidth]{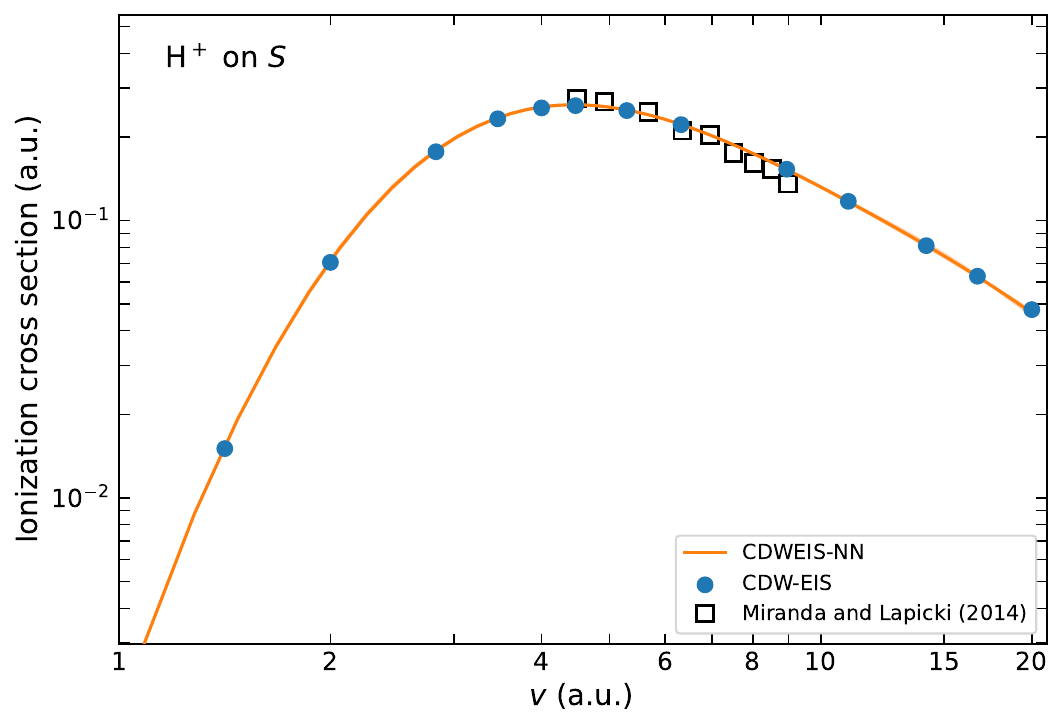}
\includegraphics[width=0.47\textwidth]{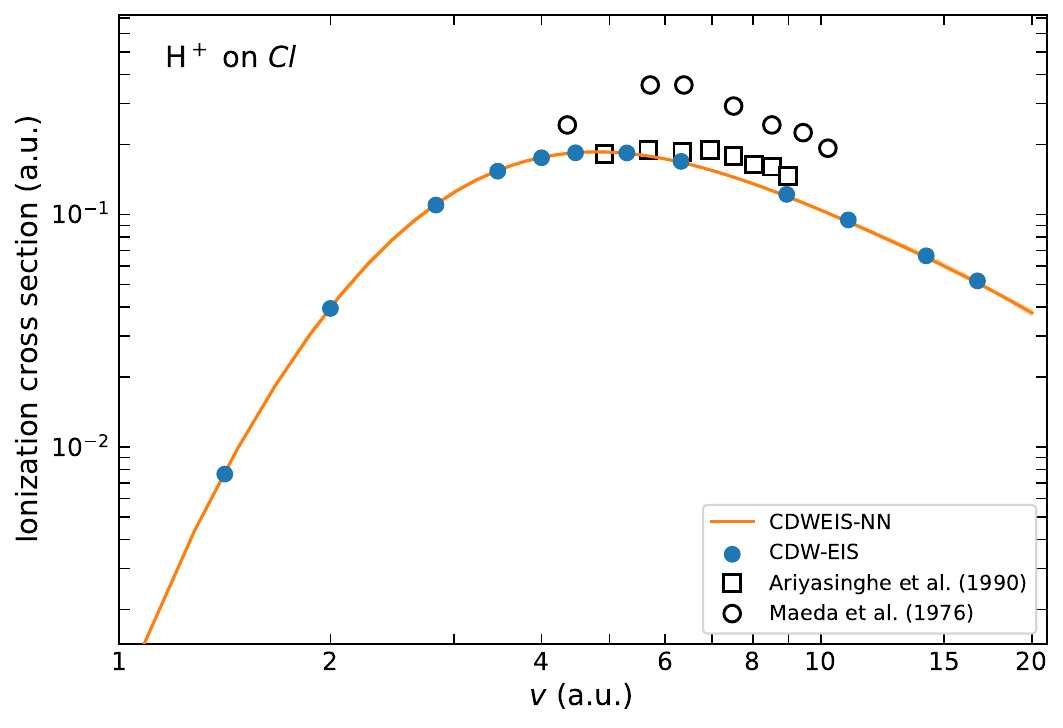}\\
\includegraphics[width=0.47\textwidth]{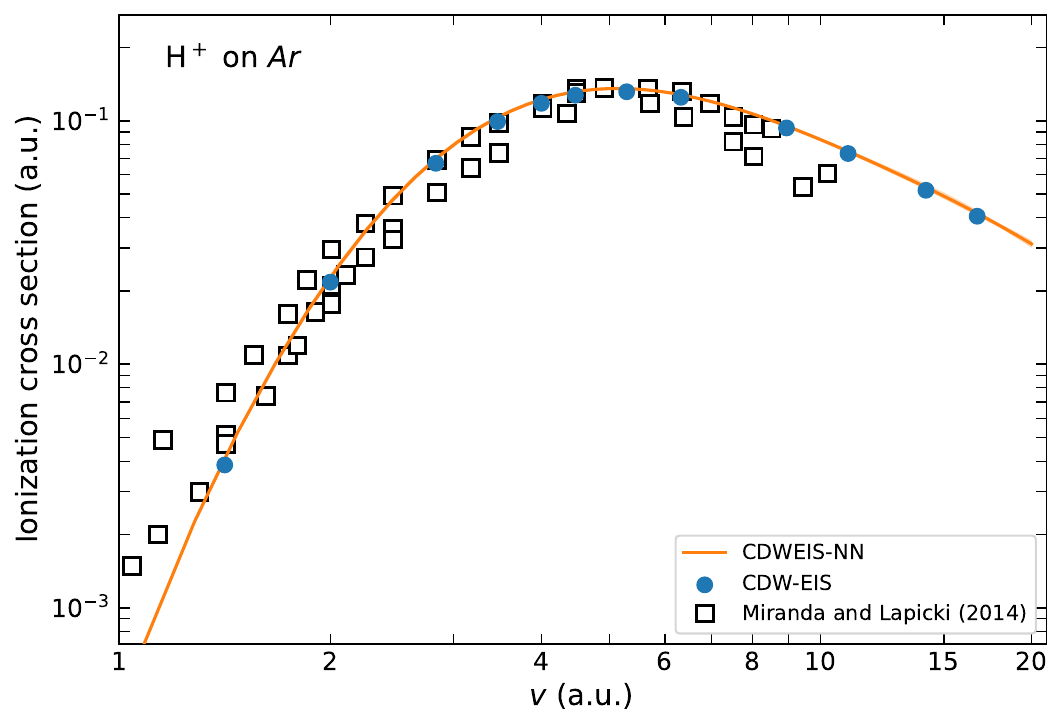}
\includegraphics[width=0.47\textwidth]{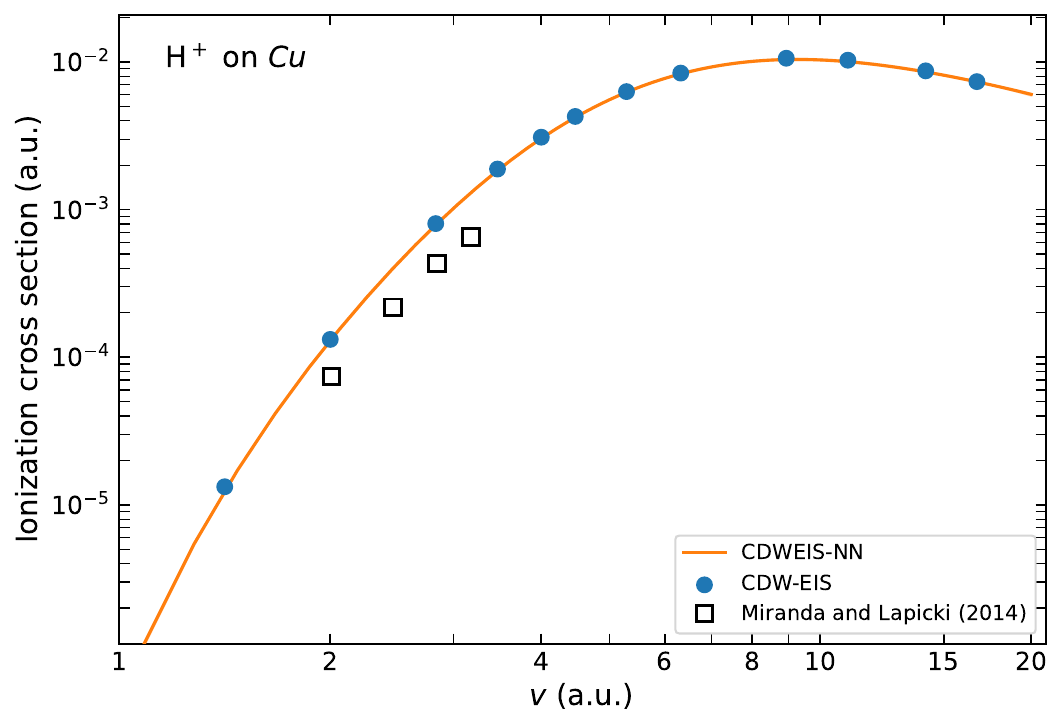}
\caption{L-shell ionization cross sections for proton impact. Filled symbols denote the CDW-EIS calculations; lines and shaded bands denote the CDWEIS-NN mean predictions and the 16th--84th percentile range of the ensemble members, respectively. Open symbols denote the experimental data compiled in Ref.~\cite{MIRANDA2014}. For Cl, squares and circles show the data of Ariyasinghe \textit{et al.} (1990)~\cite{ARIY90} and Maeda \textit{et al.} (1976)~\cite{MAE76}, respectively.}
\label{fig:P-Lshell}
\end{figure}

The subshell-resolved character of the database permits more detailed tests. Figure~\ref{fig:P-Lshell} compares proton-impact L-shell ionization cross sections with the experimental values compiled by Miranda and Lapicki \cite{MIRANDA2014}. 
The latter are taken from \cite{ARIY90} for the first five atoms displayed in the figure, with additional data from 
\cite{MAE76} for Cl and Ar, from \cite{RUD74,SCH87,STO74A,VOL70} for Ar, 
and from \cite{DEN77} for Cu.

Despite the greater diversity of target structures and experimental datasets, the calculated L-shell cross sections agree closely with the compiled values from Si to Cu. The comparison covers targets with distinct L-shell binding energies and occupancies, and hence provides a more stringent test of the subshell-resolved part of the calculation than a total-cross-section comparison alone. No systematic overestimate or underestimate is apparent across the displayed targets. The residual point-to-point dispersion is comparable to the scatter among the experimental datasets represented in the compilation.

For Cl, the compiled data form two groups: the calculations agree with the measurements of Ariyasinghe \textit{et al.} (1990)~\cite{ARIY90}, whereas the older data of Maeda \textit{et al.} (1976)~\cite{MAE76} lie systematically higher. The present L-shell data correspond mostly to high projectile velocities, at which capture into projectile states can be considered negligible. 
For the targets considered here, PIXE generally relies on K or L lines, and we found no directly comparable M-shell ionization measurements. The M$\alpha$ lines arise from transitions into the M shell from the $4f$ subshell, which is not occupied.

%%%%%%%%%%%%%%%%%%%%%%%%%%%%%%%%%%%%%%%%%%%%%%%%%%%%%%%%%%%%%%%%%%%%%%%%
%%%%%%%%%%%%%%%%%%%%%%%%%%%%%%%%%%%%%%%%%%%%%%%%%%%%%%%%%%%%%%%%%%%%%%%%
\section{The CDWEIS Neural Network Model}
\label{sec:cdwNN}

The CDW-EIS calculations presented in this work provide a physically motivated database of subshell-resolved ionization cross sections. Their numerical cost, however, prevents their direct application to every projectile--target combination and impact velocity of potential interest. We therefore developed a neural-network surrogate, denoted CDWEIS-NN, to provide a fast representation of the systematic behavior contained in the calculated database.

The present development builds on our previous experience with the IKEBANA
neural-network model, which successfully reproduced K-shell ionization cross
sections induced by electron impact~\cite{Mitnik_Atoms2025}. 
CDWEIS-NN extends this strategy to ion--atom collisions, including the dependence on the projectile charge, target structure, occupied subshell, and impact velocity.

The purpose of CDWEIS-NN is not to replace the underlying collision model. Rather, it extends the practical usefulness of the available CDW-EIS results by learning their systematic behavior within the physical domain covered by the database and by predicting projectile--target combinations that were not explicitly calculated.

%%%%%%%%%%%%%%%%%%%%%%%%%%%%%%%%%%%%%%%%%%%%%%%%%%%%%%%%%%%%%%%%%%%%%%%%%
\subsection{Training dataset and input representation}

The complete CDW-EIS repository includes subshell-resolved single-ionization cross sections $\sigma_{nl}$ for a broader range of projectile charges and target atoms than that used to train the present neural-network model. The CDWEIS-NN domain was restricted to antiprotons and bare ions with $|Z_\mathrm{P}|\leq8$, and to targets from H to Kr, $1\leq Z_\mathrm{T}\leq36$. Consequently, projectile--target systems available in the complete CDW-EIS repository but lying outside this range were not included in the training set.

This restriction defines the physical domain over which the present neural network is intended to be applied; extrapolation beyond it is not warranted. In addition, approximately forty entries at low velocities were excluded because they do not follow the smooth velocity dependence of the neighboring points (Section~\ref{subsec:applic}). 
Removing all points below a cutoff such as $v=1.5$~a.u. also worsened the agreement with the CDW-EIS calculations at higher energies, suggesting that these points help the network learn the overall velocity dependence. The remaining low-velocity results were therefore retained.

After these selections, the final training dataset comprises 6089 entries distributed over 119 distinct $(Z_\mathrm{P},Z_\mathrm{T})$ pairs.

We introduce the reduced projectile velocity 
\begin{equation}
    u=\frac{v}{\sqrt{2I_{nl}}} \, ,
\label{eq:u}
\end{equation}
which represents the ratio of the collision velocity to the mean orbital speed of the active electron (the Bohr speed of the shell).

For each sample, the target space is defined by the physical features

\begin{equation}
\mathbf{x} = \bigl(q,\, a_{\bar p},\,Z_\mathrm{T},\, n,\, l, \, N_{nl},\, \log I_{nl},\, \log v,\, \log u \bigr) \, ,
\label{eq:features}
\end{equation}
where $q=|Z_\mathrm{P}|$ is the projectile charge magnitude, $a_{\bar p}$ is an antiproton flag ($1$ for $\bar p$, $0$ otherwise). 
Separating the charge magnitude $q$ from the sign of the projectile charge via this flag allows a single network to represent both proton and antiproton collisions, capturing the physically distinct dynamics that arise from the sign of 
$Z_\mathrm{P}$ at low velocities \cite{Kirchner_2011, Montanari_2012} without requiring two independent models.
$N_{nl}$ is the electron occupation of the ionized shell, and $I_{nl}$ and $u$ have been defined in Eqs.~(\ref{eq:Inl}) and (\ref{eq:u}). The logarithmic scaling of energy and velocity is introduced to compress the dynamic range of the input space. Although $\log u$ and $\log v$ are not independent given $\log I_{nl}$, both are retained in the input: $\log u$ encodes the collision regime relative to the ionization threshold, while $\log v$ enters the collision dynamics independently through the eikonal factor $\xi_v = Z_P/v$; the dependence on $Z_P$ is accounted for separately through $q$, allowing the network to construct the relevant combination $v/Z_P$ implicitly.  
All input variables are standardized independently using the statistics of the corresponding training partition. 

The regression target is the logarithm of the ionization cross section per electron,
\begin{equation}
y=\log\left(\frac{\sigma_{nl}}{N_{nl}}\right).
\label{eq:target}
\end{equation}
The logarithmic transformation is required because the cross sections span approximately 27 orders of magnitude. Dividing by $N_{nl}$ removes the trivial contribution associated with the number of equivalent electrons and produces a smoother target distribution. The physical subshell cross section is recovered from the predicted value $\hat y$ through
\begin{equation}
\hat{\sigma}_{nl}=N_{nl}\,10^{\hat y}.
\label{eq:sigmapred}
\end{equation}

%%%%%%%%%%%%%%%%%%%%%%%%%%%%%%%%%%%%%%%%%%%%%%%%%%%%%%%%%%%%%%%%%%%%%%%%%
\subsection{Architecture, training, and final ensemble}
\label{subsec:Archit}

A conventional division into fixed training, validation, and test sets would reserve a substantial fraction of the present database for a single final test. Because the dataset is relatively small, spans many orders of magnitude, and has an irregular coverage of projectile--target pairs, this would leave some regions of the physical domain poorly represented during training. We therefore adopted a repeated group-held-out ensemble strategy.

The CDWEIS-NN is implemented as an ensemble of $M=5$ feed-forward neural networks trained independently. Each network has an input layer of dimension 9 [Eq.~\eqref{eq:features}], three fully connected hidden layers with $(64,64,32)$ neurons, and a single linear output neuron, totaling approximately 6900 trainable parameters. Hyperbolic-tangent activation functions are used throughout the hidden layers. All weight matrices are initialized with the Xavier uniform scheme, and all bias vectors are initialized to zero. The only explicit regularization employed is $L_2$ weight decay; no dropout is used.

To avoid information leakage, the data are divided into complete projectile--target groups, $g=(Z_\mathrm{P},Z_\mathrm{T})$. A random point-by-point split would allow different velocities or subshells of the same collision system to appear in both training and validation data, leading to an overly optimistic estimate of predictive performance. Instead, each ensemble member is trained on a random 90\% of the projectile--target pairs, while the remaining 10\% of complete pairs are reserved for validation and early stopping. The five members therefore differ both in their initialization and in the systems used for training.

The networks are optimized with Adam, using a learning rate of $8\times10^{-4}$, mini-batches of 128 samples, and an $L_2$ weight decay of $10^{-4}$. The loss function is the mean-squared error in the logarithmic target of Eq.~\eqref{eq:target}. The learning rate is reduced by a factor of two if the validation loss does not improve over 50 consecutive validation checks. Validation is performed every five epochs, and training is stopped after 80 consecutive checks without improvement, with a maximum of 2000 epochs. The parameters corresponding to the minimum validation loss are retained.

\begin{figure}[H]
    \centering
    \includegraphics[width=\textwidth]{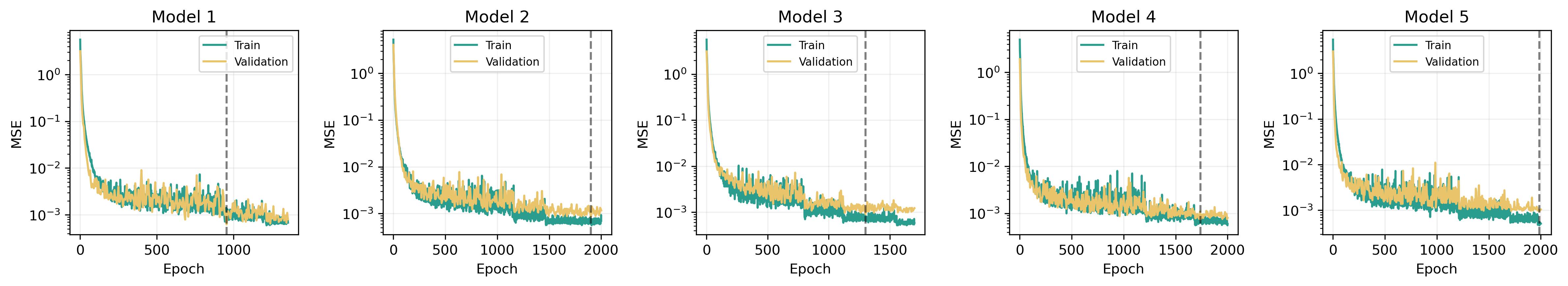}
    \caption{Evolution of the mean-squared-error loss in the logarithmic target during training of the five CDWEIS-NN ensemble members. Solid and dashed curves denote the training and group-held-out validation losses, respectively. The vertical dashed line marks the epoch of the retained checkpoint, corresponding to the minimum validation loss, including runs that reached the maximum of 2000 epochs. The similar evolution of the five training--validation pairs indicates stable optimization and no evident validation-loss divergence during training.}
    \label{fig:learning-curves}
\end{figure}

Figure~\ref{fig:learning-curves} shows the evolution of the training and validation losses for the five ensemble members. In all cases, both losses decrease rapidly during the initial epochs and subsequently evolve at similar values without a sustained divergence. Some members satisfy the early-stopping criterion before reaching the maximum number of epochs, whereas the remaining runs reach the prescribed limit of 2000 epochs. The consistency of the five learning curves supports the stability of the optimization and model-selection procedure.

For a given input $\mathbf{x}$, each network $m$ produces a prediction $y^{(m)}(\mathbf{x})$. The ensemble statistics are
\begin{equation}
\bar y(\mathbf{x}) = \frac{1}{M}\sum_{m=1}^{M} y^{(m)}(\mathbf{x}),\qquad
\sigma_{\rm ens}(\mathbf{x}) = \sqrt{\frac{1}{M}\sum_{m=1}^{M}\bigl[y^{(m)}-\bar y\bigr]^2} \, .
\label{eq:ensemble}
\end{equation}
This averaging reduces the dependence on a particular random initialization and on the specific training subset used for each member. The 16th--84th percentile range of the member predictions is shown in selected figures as an indication of model-to-model spread. It provides a useful local diagnostic: a narrow range indicates close agreement among the five models, whereas a broader range identifies regions where their predictions differ more strongly. This spread does not include all sources of predictive uncertainty, such as the discrepancy of the underlying CDW-EIS model with experiment; it is therefore interpreted as an ensemble-agreement indicator rather than as a fully calibrated confidence interval.

%%%%%%%%%%%%%%%%%%%%%%%%%%%%%%%%%%%%%%%%%%%%%%%%%%%%%%%%%%%%%%%%%%%%%%%%%
\subsection{Repeated group-held-out validation}
\label{subsec:heldout}

The five group-preserving splits provide five independent opportunities to test the model on projectile--target pairs that were not used to train a given ensemble member. For each member, the cross sections of its validation pairs are predicted after training and compared with the corresponding CDW-EIS values.

The validation predictions from the five members are then collected into a single held-out dataset. If a projectile--target pair was selected for validation in only one split, its cross sections contribute the predictions of that member. If the same pair was selected in two or more splits, its predictions are averaged using only the members for which that pair was absent from training. In no case is a prediction from a member trained with the same $(Z_\mathrm{P},Z_\mathrm{T})$ pair included in the held-out evaluation.

The validation strategy employed here differs from a conventional $k$-fold cross-validation in an important respect: the five group-preserving splits are drawn independently at random and are therefore not constrained to be mutually exclusive. The five validation selections may thus overlap, meaning that some projectile–target pairs could be held out in more than one split while others may not be held out at all. This design was chosen to maximize the training data available to each member while still evaluating generalization to unseen collision systems.
In the present case, 47 of the 119 projectile--target pairs were selected for validation at least once. They comprise 2371 cross sections, corresponding to 38.9\% of the filtered dataset. The resulting set provides a broad sample of predictions for previously unseen collision systems.

We assess the model fidelity through three complementary metrics:

\begin{enumerate}
    \item Log-space absolute error:
    \begin{equation}
        \varepsilon_{\log} =
        \left|
        \log \sigma_{\rm true}
        -
        \log \sigma_{\rm pred}
        \right| \, .
    \end{equation}
    Its average over a set of predictions defines the log-space mean
    absolute error (MAE). This is the natural measure for the logarithmic
    regression target and weights relative deviations uniformly across the
    full dynamic range of the cross sections. An error
    $\varepsilon_{\log}=\delta$ corresponds to a multiplicative deviation
    of $10^\delta$; for example, $\varepsilon_{\log}=0.02$ corresponds to
    a factor $10^{0.02}\approx1.05$.

    \item Relative error:
    \begin{equation}
        \varepsilon_{\rm rel} =
        \left|
        \frac{\sigma_{\rm pred}-\sigma_{\rm true}}
        {\sigma_{\rm true}}
        \right|
        \times100\% \, .
    \end{equation}
    It expresses the discrepancy directly as a percentage and can therefore
    be compared readily with experimental uncertainties.

    \item Symmetric error factor:
    \begin{equation}
        f =
        \max\left(
        \frac{\sigma_{\rm pred}}{\sigma_{\rm true}},
        \frac{\sigma_{\rm true}}{\sigma_{\rm pred}}
        \right) \, .
    \end{equation}
    This quantity measures the multiplicative discrepancy without regard to
    its sign. For example, $f=2$ means that the prediction differs from the
    reference value by at most a factor of two.
\end{enumerate}

Table~\ref{tab:nn-metrics} summarizes the resulting held-out performance. The log-space MAE is 0.0192, corresponding to a characteristic multiplicative deviation of $10^{0.0192}\simeq1.045$. The median relative error is 3.1\%; 90\% of the predictions have relative errors below 9.2\%, and 99.5\% differ by less than 30\% from the corresponding CDW-EIS value. The MAEs obtained separately for the five members range from 0.0182 to 0.0217, showing that the result is not driven by one particular training split.

\begin{table}[H]
\caption{Performance of CDWEIS-NN for repeated group-held-out projectile--target pairs. For every point, only predictions from ensemble members that did not use its $(Z_\mathrm{P},Z_\mathrm{T})$ pair during training are included.}
\label{tab:nn-metrics}
\centering
\begin{tabular}{lcc}
\toprule
\textbf{Metric} & \textbf{Mean} & \textbf{Median} \\
\midrule
Log-space absolute error $\varepsilon_{\log}$ & 0.019 & 0.013 \\
Relative error $\varepsilon_{\rm rel}$ & 4.41\% & 3.06\% \\
Error factor $f$ & 1.047 & 1.031 \\
\bottomrule
\end{tabular}
\end{table}

The held-out dataset comprises 2371 cross sections from 47 of the 119 projectile--target pairs. In addition, 91.5\%, 98.2\%, and 99.5\% of the held-out predictions lie within 10\%, 20\%, and 30\%, respectively, of the corresponding CDW-EIS value.

The validation pairs are also used to determine the stopping epoch of their corresponding models. The present procedure should therefore be understood as repeated group-held-out validation rather than as a completely independent final test. Nevertheless, it directly evaluates the intended application of CDWEIS-NN: predicting the cross sections of projectile--target pairs absent from the training data.

%%%%%%%%%%%%%%%%%%%%%%%%%%%%%%%%%%%%%%%%%%%%%%%%%%%%%%%%%%%%%%%%%%%%%%%%%
\subsection{Reconstruction by the final ensemble}

The repeated held-out results above assess the predictive capability of the individual ensemble members. Once the five members have been trained, they are combined through Eq.~\eqref{eq:ensemble} to produce the final CDWEIS-NN emulator used in the remainder of this work.

The complete set of performance metrics is summarized in Table~\ref{tab:nn-globalmetrics}. 

\begin{table}[H]
\caption{Global performance metrics of the CDWEIS-NN ensemble on the full filtered dataset.\label{tab:nn-globalmetrics}}
\centering
\begin{tabular}{lcc}
\toprule
\textbf{Metric} & \textbf{Mean} & \textbf{Median} \\
\midrule
Log-space absolute error $\varepsilon_{\log}$ & 0.013 & 0.009 \\
Relative error $\varepsilon_{\rm rel}$ & 3.08\% & 2.10\% \\
Error factor $f$  & 1.032 & 1.021 \\
\bottomrule
\end{tabular}
\end{table}

\begin{figure}[H]
\centering
\includegraphics[width=0.32\textwidth]{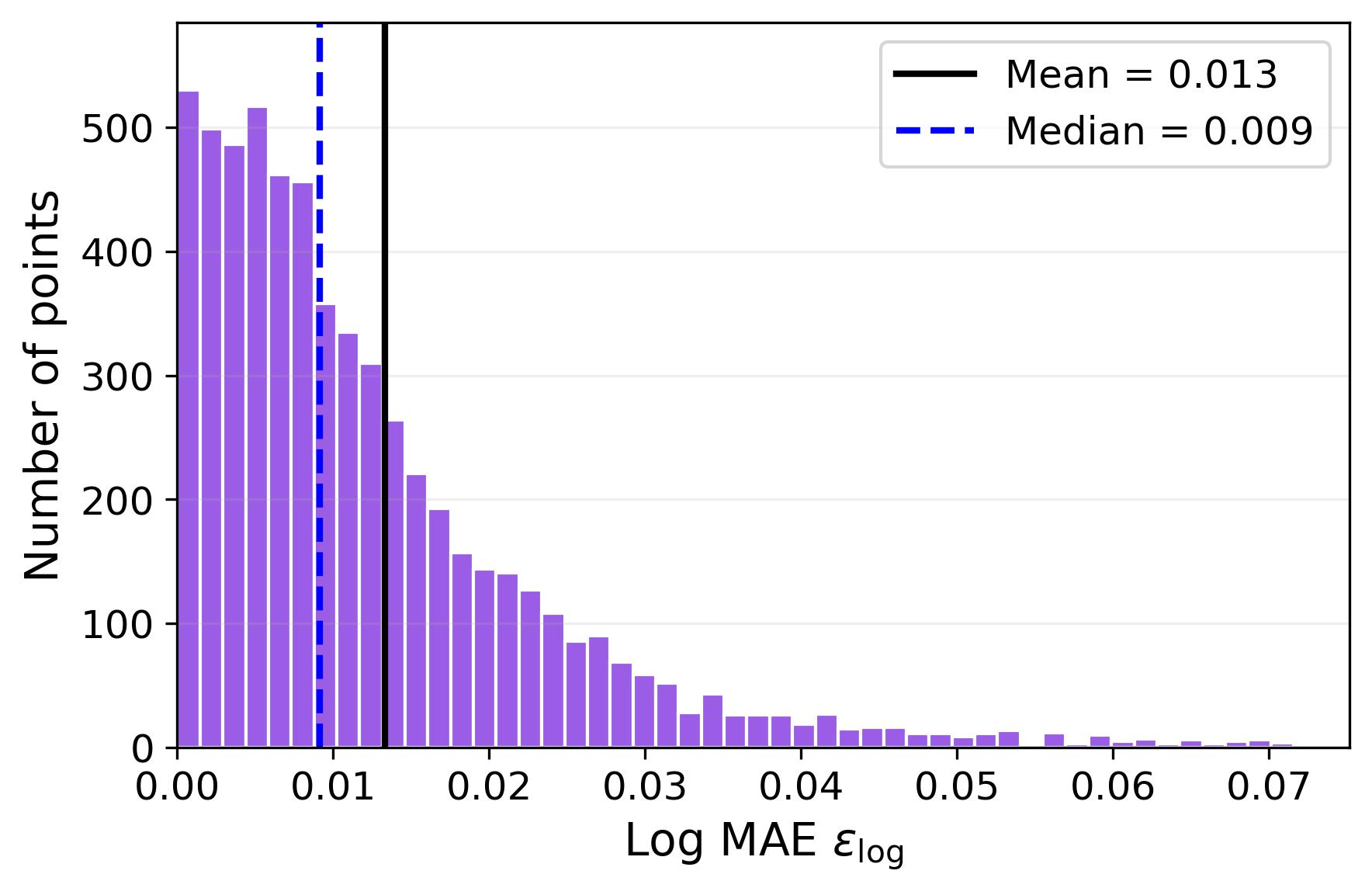}
\includegraphics[width=0.32\textwidth]{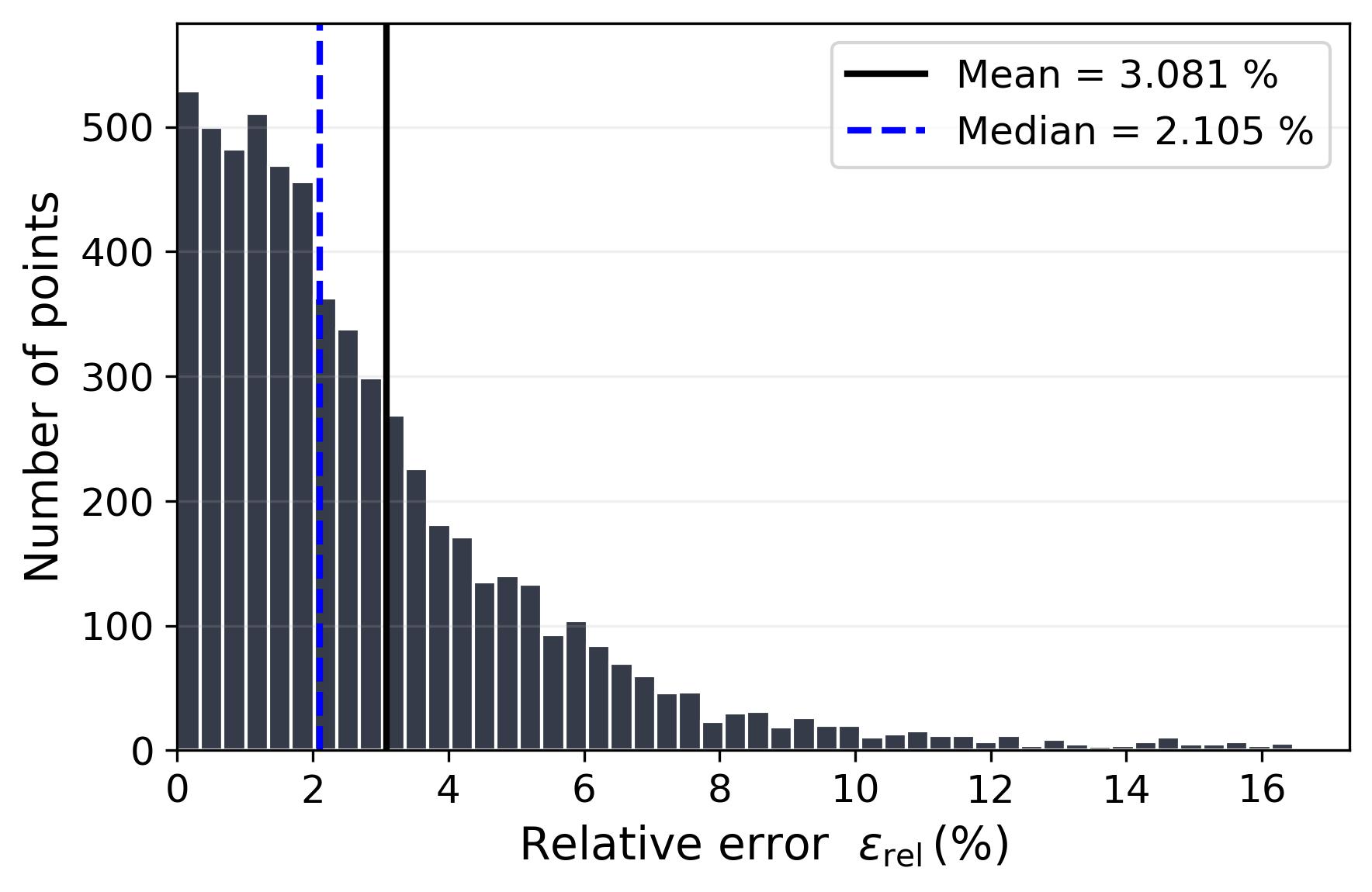}
\includegraphics[width=0.32\textwidth]{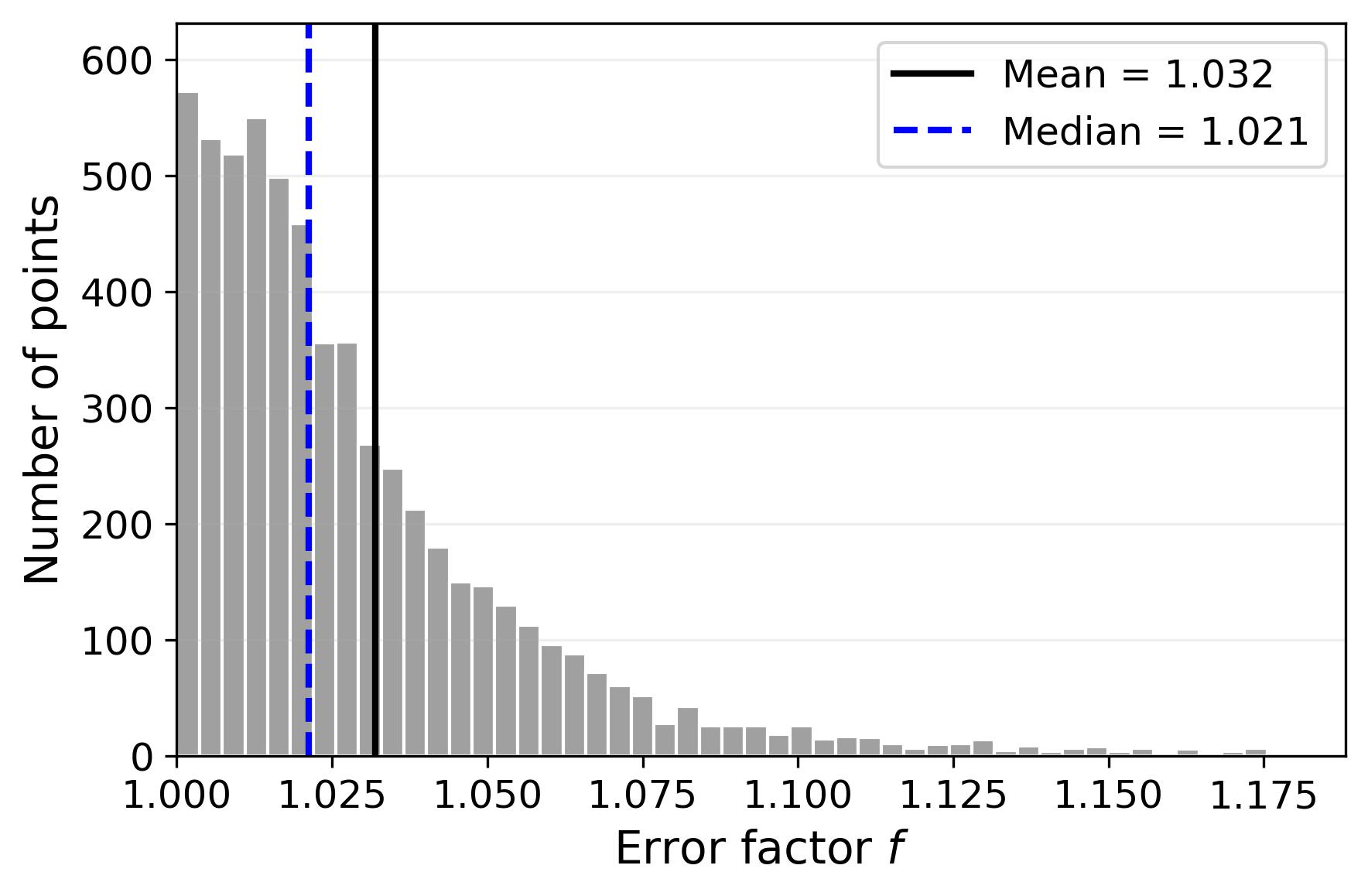}
\caption{Error distributions for the final five-member CDWEIS-NN ensemble evaluated on the complete filtered CDW-EIS database. Left: absolute logarithmic error $\varepsilon_{\log}$. Center: relative error $\varepsilon_{\rm rel}$. Right: symmetric error factor $f$. Vertical lines indicate the corresponding mean and median values.}
\label{fig:final-ensemble-histograms}
\end{figure}

\begin{figure}[H]
\centering
\includegraphics[width=0.60\textwidth]{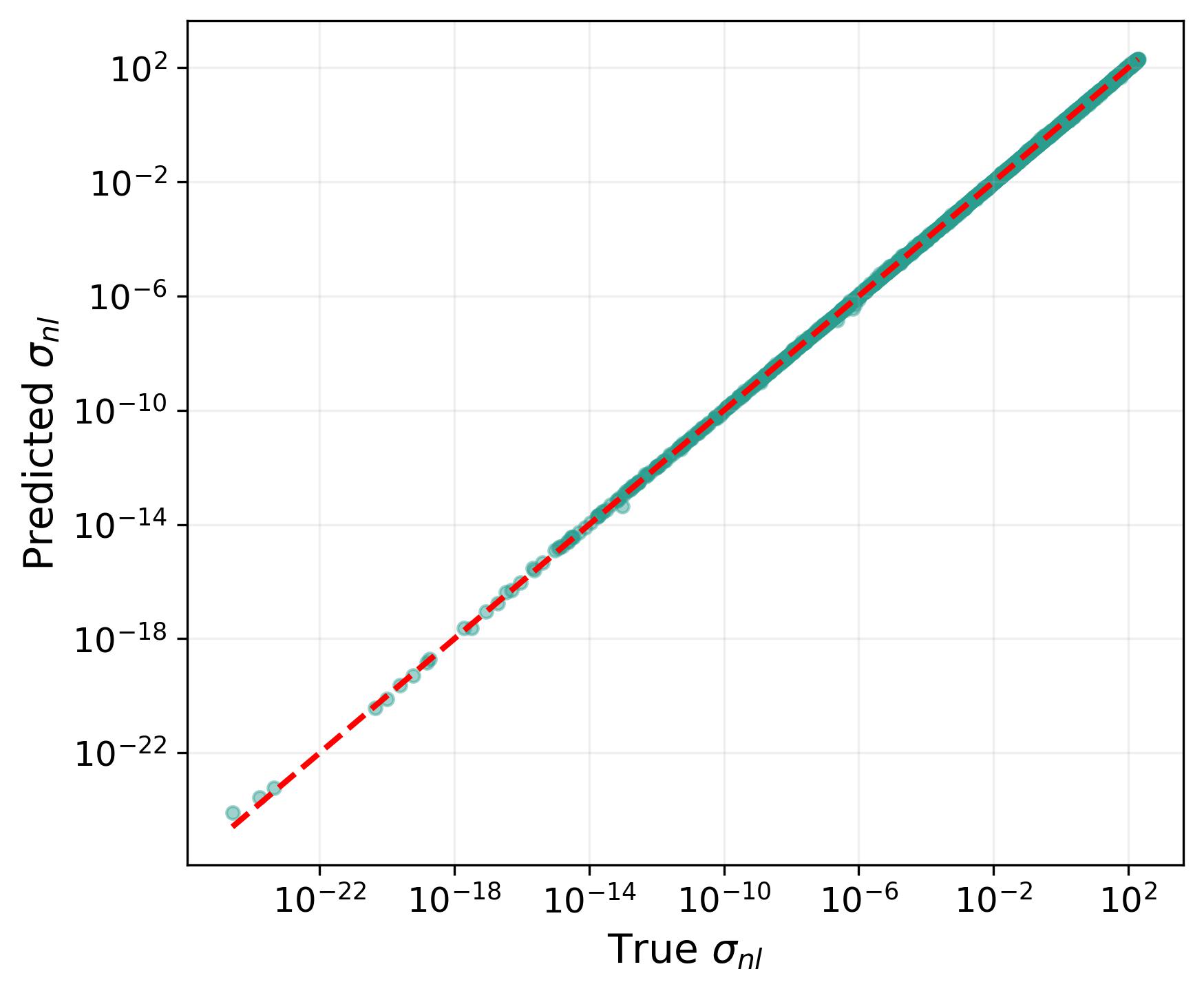}
\caption{Predicted versus calculated values of $\log(\sigma_{nl}/N_{nl})$ for the complete filtered CDW-EIS database. The diagonal line indicates equality between the CDWEIS-NN prediction and the CDW-EIS reference.}
\label{fig:pred-vs-real}
\end{figure}

Figure~\ref{fig:final-ensemble-histograms} shows the distributions of the three error measures for the final ensemble, while Figure~\ref{fig:pred-vs-real} compares its predictions with the CDW-EIS cross sections across the full filtered dataset.
The narrow error distributions and the close agreement between predicted and calculated values demonstrate that CDWEIS-NN accurately and consistently reproduces the CDW-EIS results over their broad dynamic range.


The global reconstruction and repeated held-out results tell a consistent story. The held-out errors are moderately larger, as expected for previously unseen collision systems, but remain small across the tested domain. Together, the two analyses show that the five individual models perform consistently and that their average provides a stable representation of the CDW-EIS database.

%%%%%%%%%%%%%%%%%%%%%%%%%%%%%%%%%%%%%%%%%%%%%%%%%%%%%%%%%%%%%%%%%%%%%%%%%
\subsection{Representative CDWEIS-NN predictions}

\begin{figure}[H]
\centering
\includegraphics[width=0.48\textwidth]{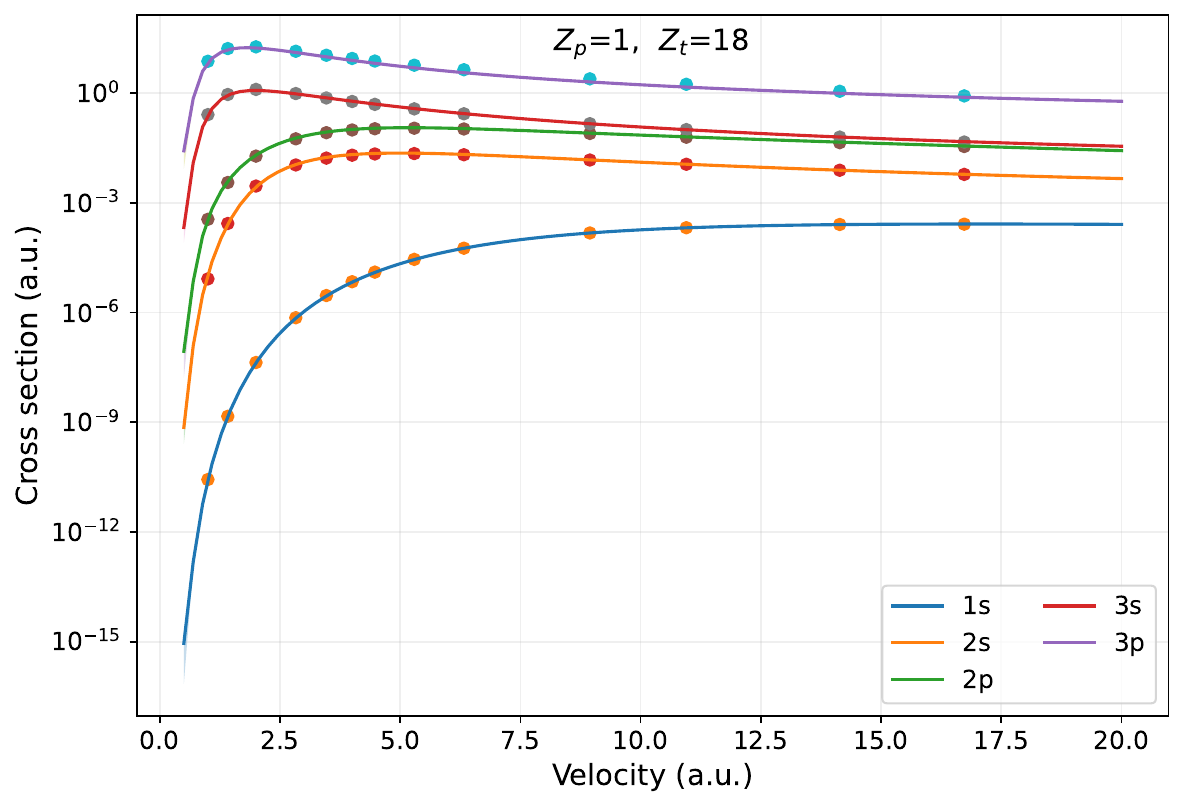}
\includegraphics[width=0.48\textwidth]{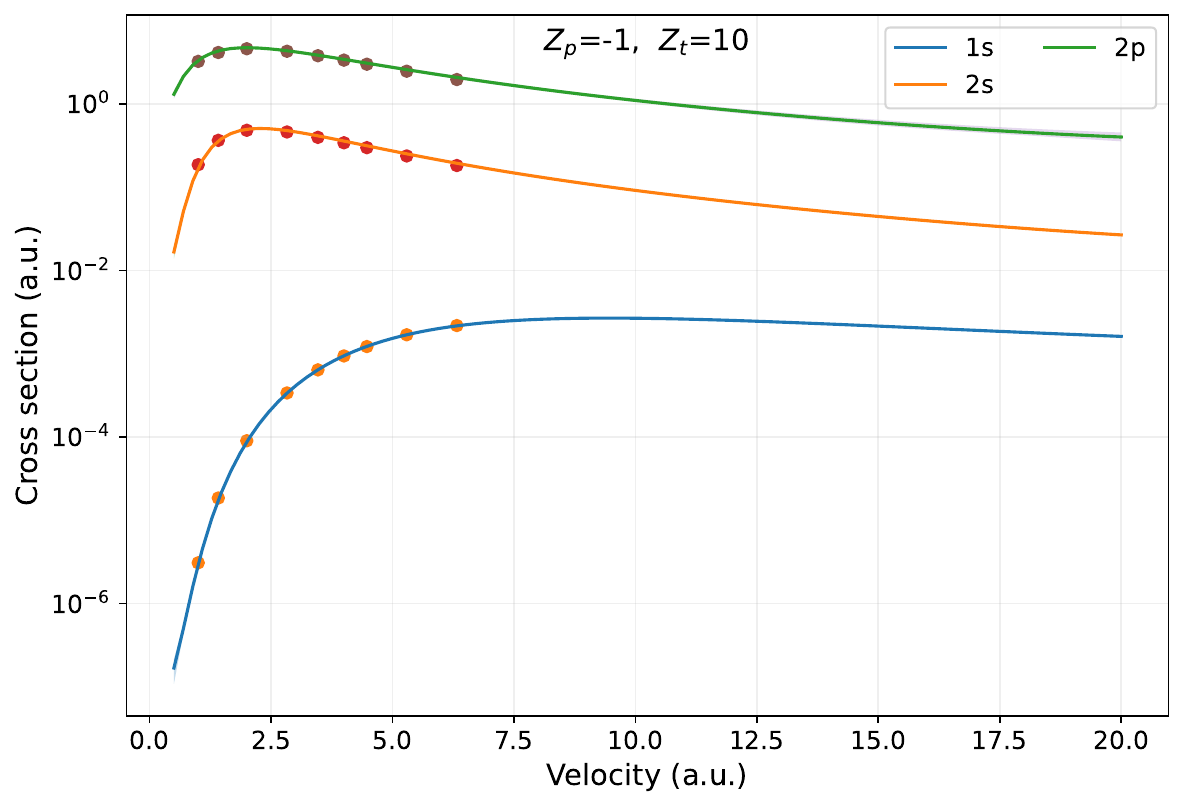} \\
\includegraphics[width=0.48\textwidth]{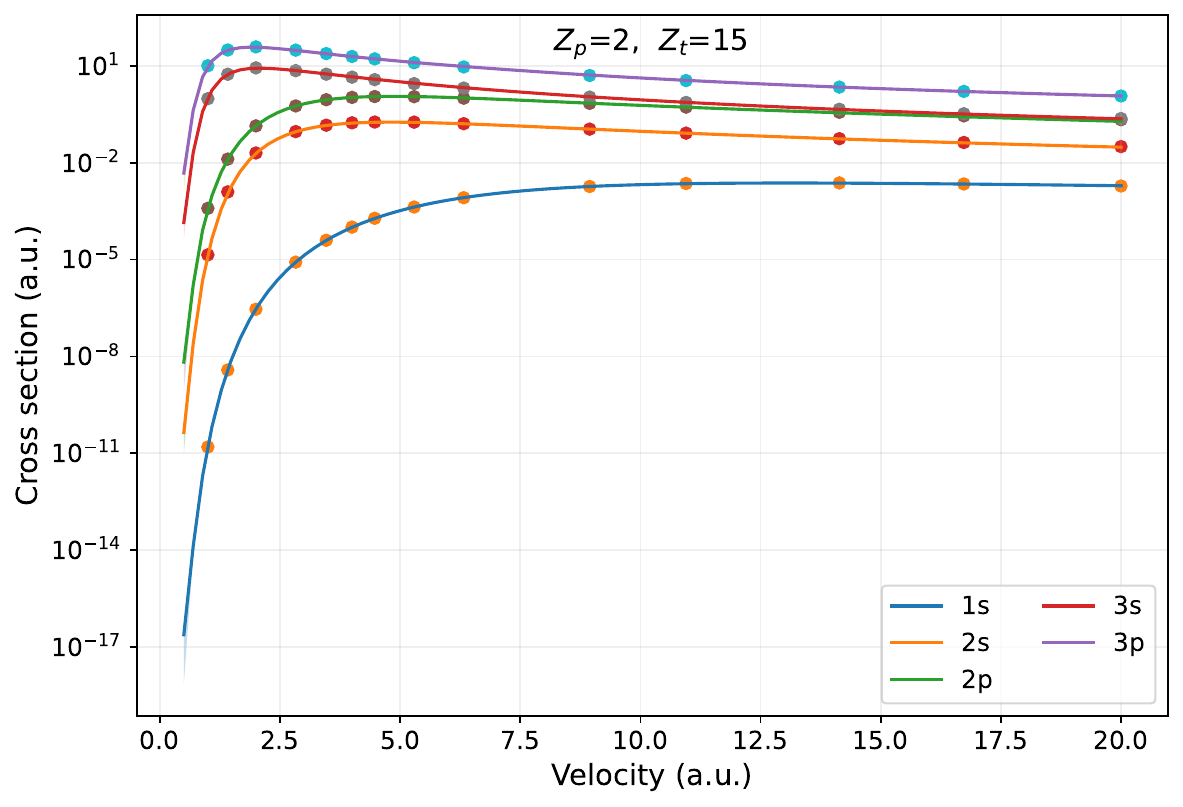}
\includegraphics[width=0.48\textwidth]{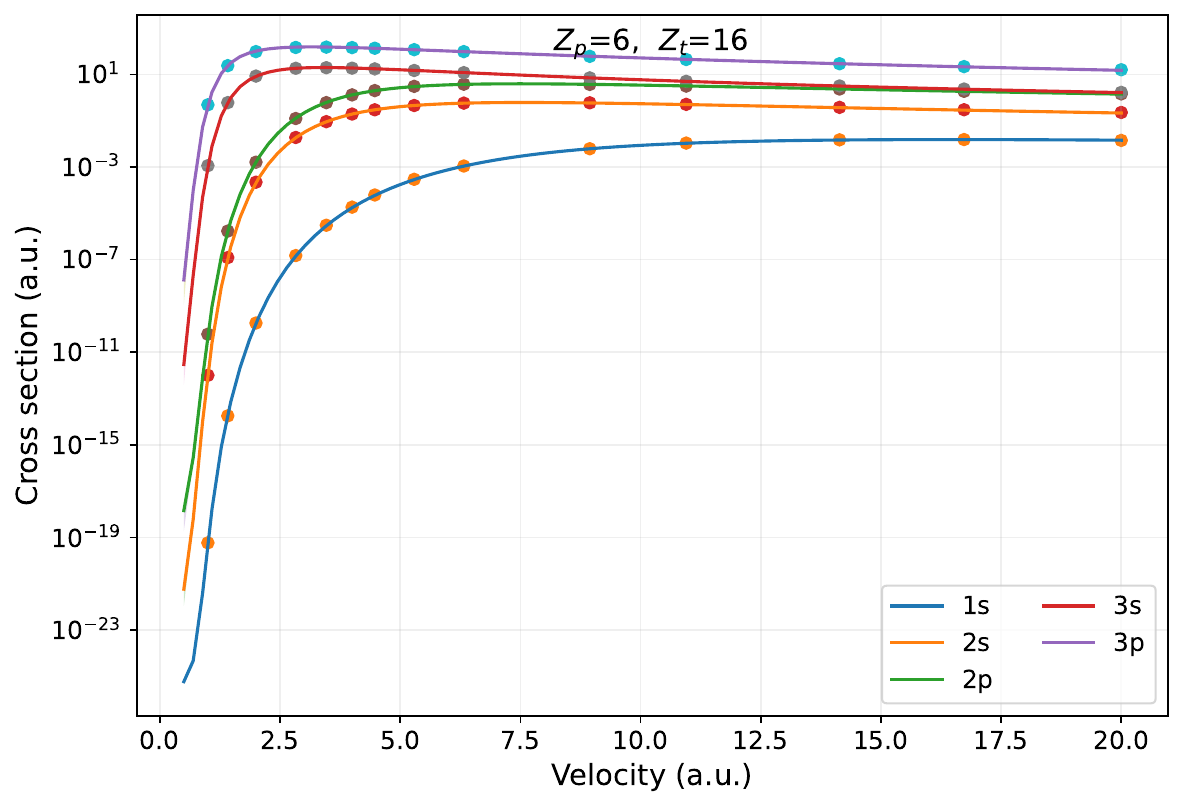}
\caption{Subshell-resolved cross sections predicted by CDWEIS-NN for proton impact on Ar (upper left), antiproton impact on Ne (upper right), He$^{2+}$ impact on P (lower left), and C$^{6+}$ impact on S (lower right). The lines denote the final ensemble prediction.
The shaded ranges, which indicate the 16th–84th percentile spread of the ensemble members are barely visible on the logarithmic scale. The symbols represent CDW-EIS calculations.}
\label{fig:example-inner}
\end{figure}

Figure~\ref{fig:example-inner} presents subshell-resolved predictions for representative systems in the CDW-EIS database. The continuous curves are the final ensemble predictions evaluated on a dense velocity grid, whereas the symbols are the original CDW-EIS values. The network reproduces the ordering of the shell contributions, the low-velocity rise, the broad cross-section maxima, and the high-velocity decrease. No asymptotic forms or results from other theoretical models have been included in its training.

\begin{figure}[H]
\centering
\includegraphics[width=0.48\textwidth]{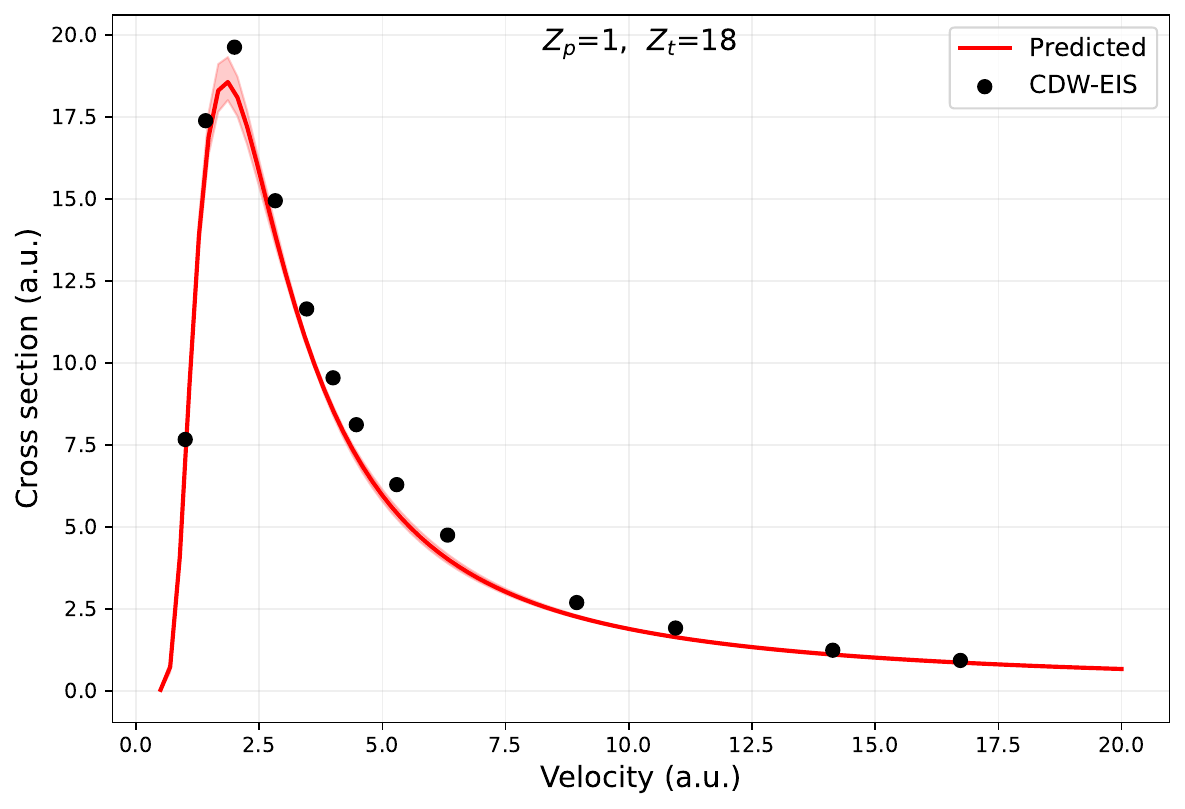}
\includegraphics[width=0.48\textwidth]{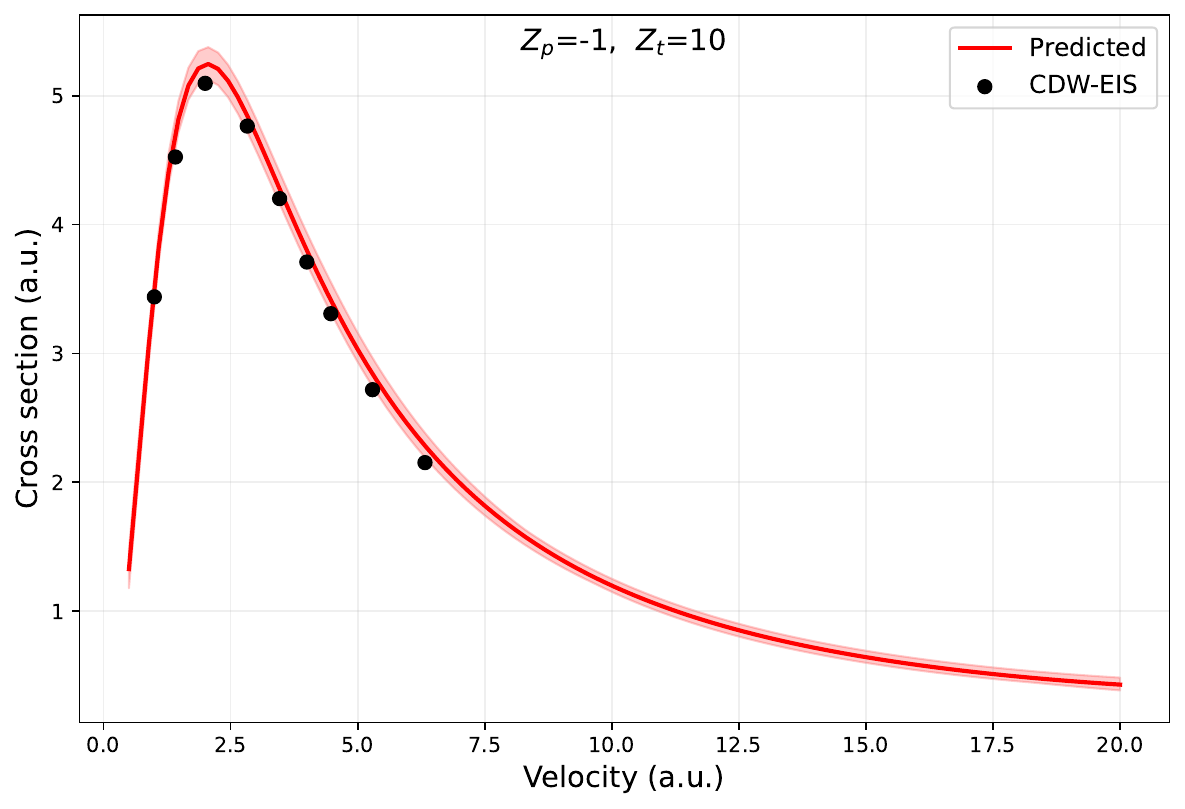} \\
\includegraphics[width=0.48\textwidth]{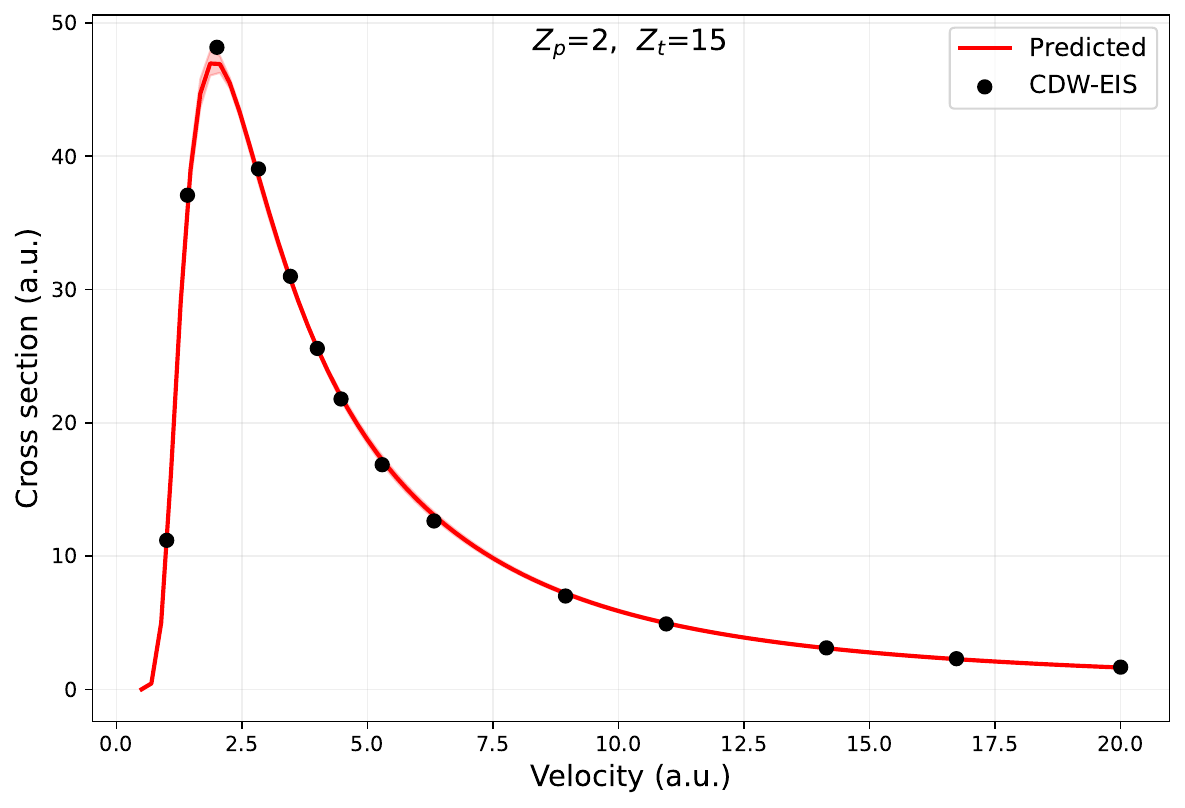}
\includegraphics[width=0.48\textwidth]{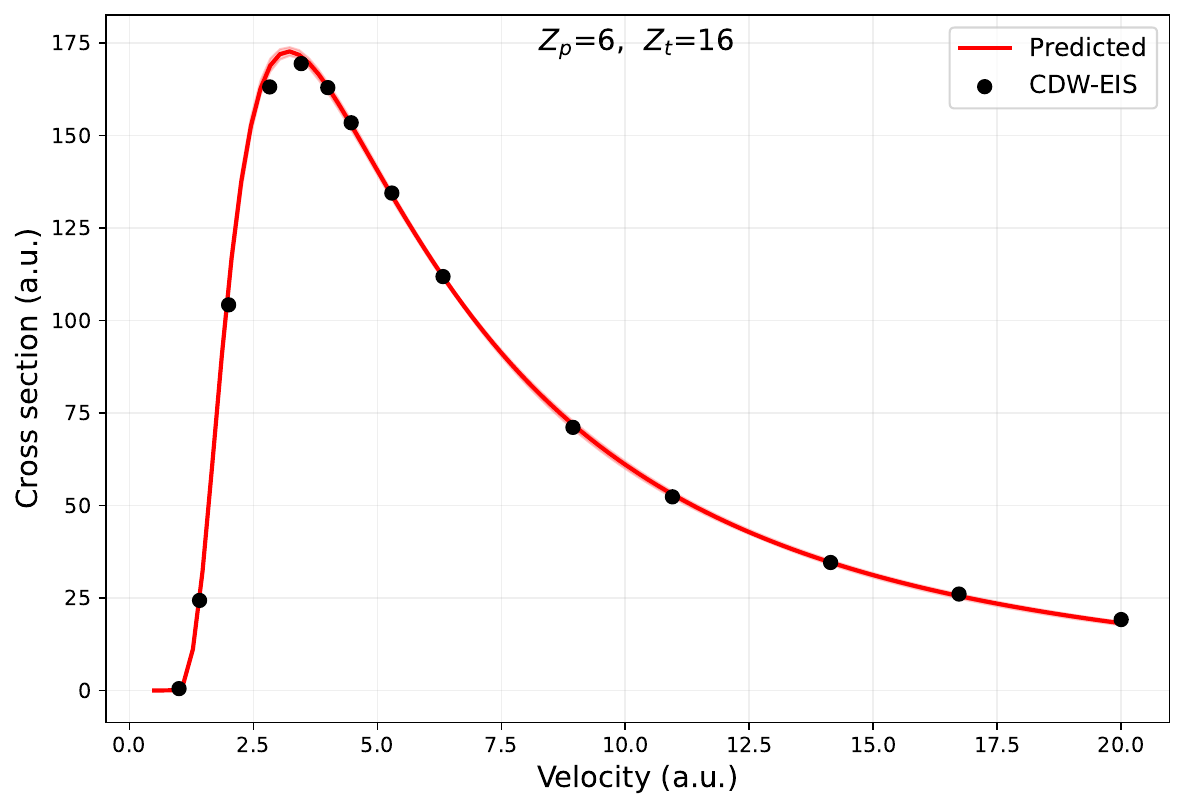}
\caption{Total ionization cross sections for the systems shown in Fig.~\ref{fig:example-inner}. Lines denote final CDWEIS-NN predictions, shaded regions indicate the 16th--84th percentile range of the ensemble members, and symbols represent CDW-EIS calculations.}
\label{fig:example-total}
\end{figure}

The corresponding total single-ionization cross sections,
\begin{equation}
\sigma_{\rm tot}(v)=\sum_{nl}\sigma_{nl}(v),
\end{equation}
are shown in Fig.~\ref{fig:example-total}. For each ensemble member, the shell contributions are summed before calculating the ensemble statistics. This procedure preserves the correlations among the partial cross sections when constructing the total ionization curve.

\begin{figure}[H]
\centering
\includegraphics[width=0.48\textwidth]{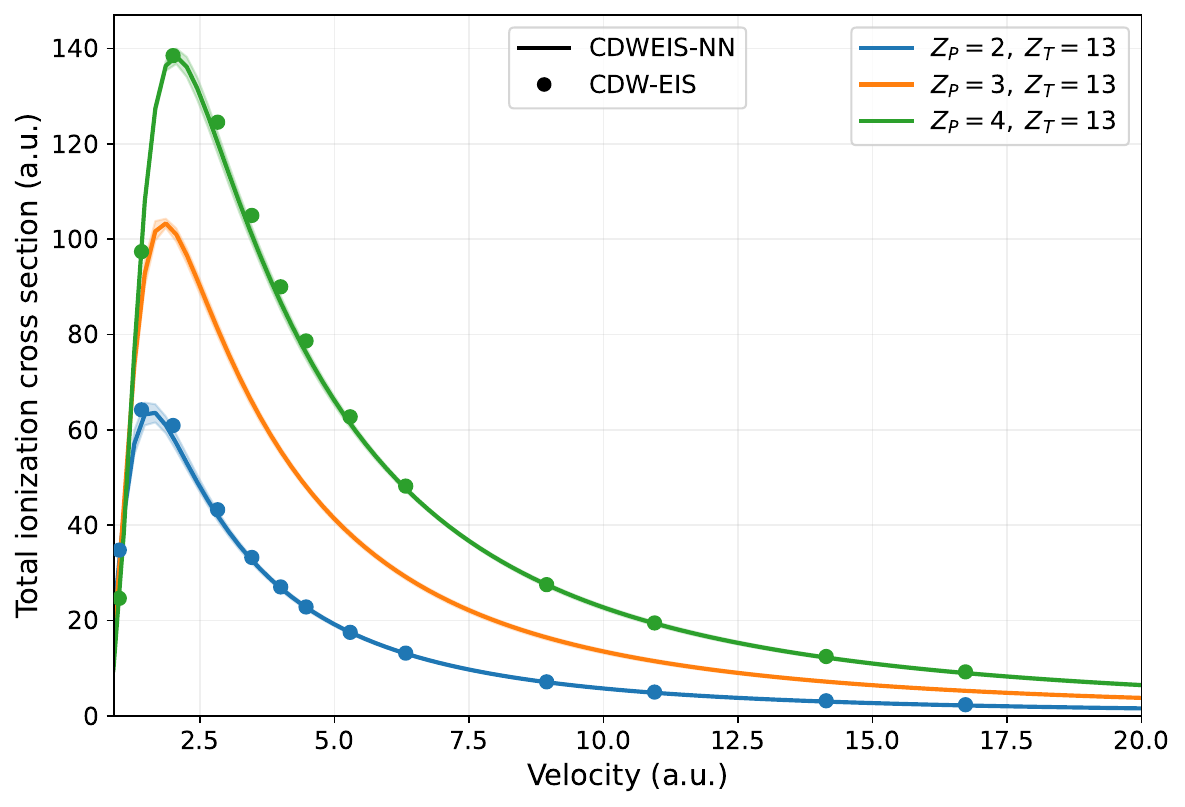}
\includegraphics[width=0.48\textwidth]{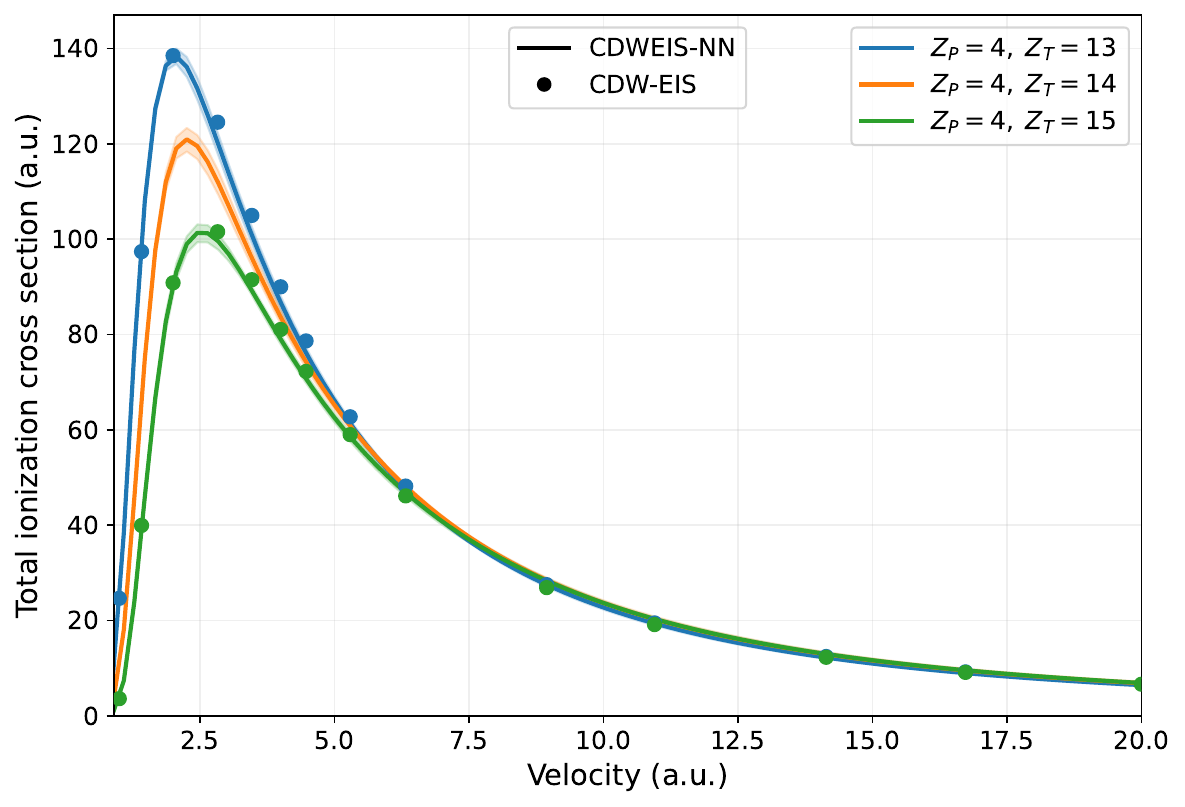}
\caption{CDWEIS-NN predictions for total ionization cross sections. Left: fixed Al target ($Z_\mathrm{T}=13$) and bare projectiles with $Z_\mathrm{P}=2$, 3, and 4. Right: fixed projectile charge $Z_\mathrm{P}=4$ and Al, Si, and P targets. The systems $(Z_\mathrm{P},Z_\mathrm{T})=(3,13)$ and $(4,14)$ were not explicitly calculated in the CDW-EIS database.}
\label{fig:compNN}
\end{figure}

The ability of the network to predict missing systems is illustrated in Fig.~\ref{fig:compNN}. The left panel shows total ionization cross sections for a fixed Al target and projectile charges $Z_\mathrm{P}=2$, 3, and 4. The system $(Z_\mathrm{P},Z_\mathrm{T})=(3,13)$ was not explicitly calculated in the CDW-EIS database. The right panel considers a fixed projectile charge, $Z_\mathrm{P}=4$, and the Al, Si, and P targets; the pair $(4,14)$ is likewise absent from the calculated dataset. In both cases, the predictions follow the systematic behavior defined by the neighboring available systems.

To facilitate the use of the surrogate model, the trained CDWEIS-NN ensemble is also provided through the self-contained 
\href{https://colab.research.google.com/drive/1TrblWWwpTEwSlGh--dSyBWkIYNvUbksg}{\texttt{CDWEISNN.ipynb}} Google Colab notebook \cite{CDWEIS-NN}.
Users can run the notebook directly in a web browser, without installing local software, specify the projectile--target pair $(Z_\mathrm{P},Z_\mathrm{T})$, and obtain the predicted subshell-resolved and total ionization cross sections together with the corresponding tables and figures. 
%%%%%%%%%%%%%%%%%%%%%%%%%%%%%%%%%%%%%%%%%%%%%%%%%%%%%%%%%%%%%%%%%%%%%%%%%
\subsection{Final assessment: predictions against experiment}
\label{subsec:finalass}

The comparisons of CDWEIS-NN with held-out CDW-EIS calculations establish the ability of the network to emulate the theoretical database. The final and practically most relevant assessment is whether the resulting predictions are consistent with experimental ionization measurements for systems that were not explicitly calculated.

\begin{figure}[H]
\centering
\includegraphics[width=0.47\textwidth]{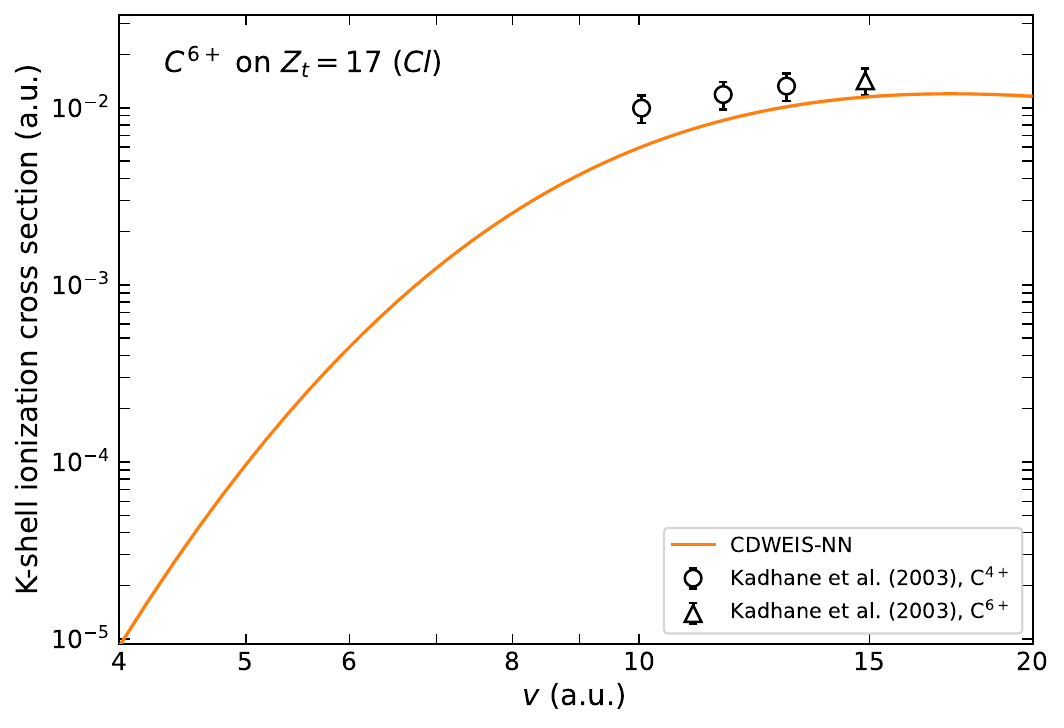}
\includegraphics[width=0.47\textwidth]{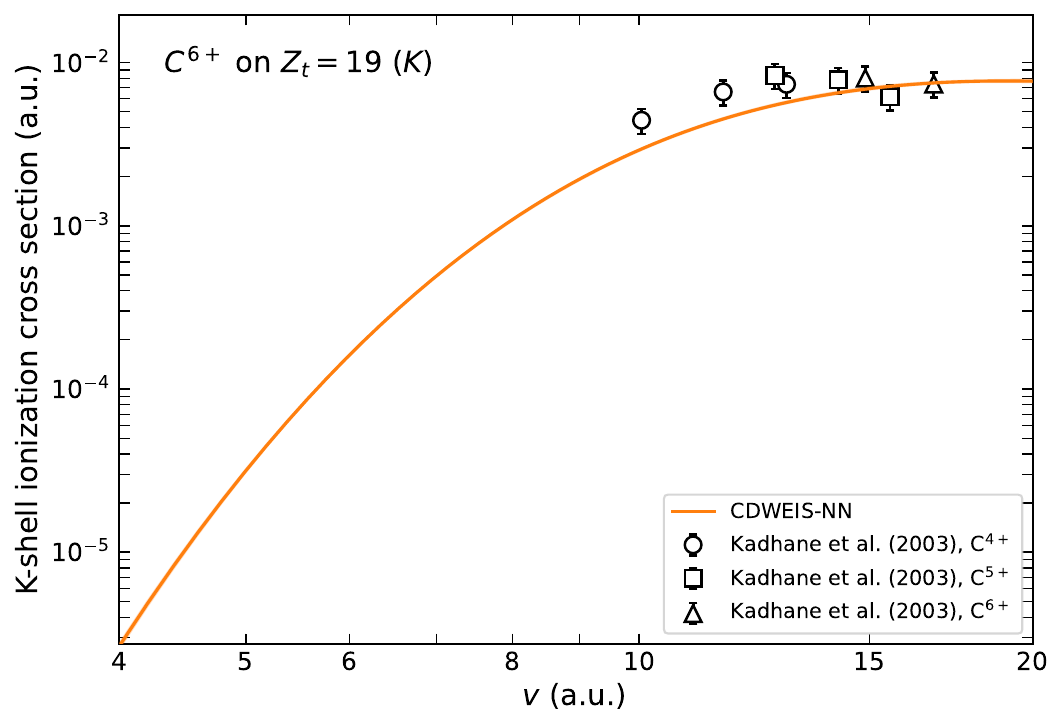}\\
\includegraphics[width=0.47\textwidth]{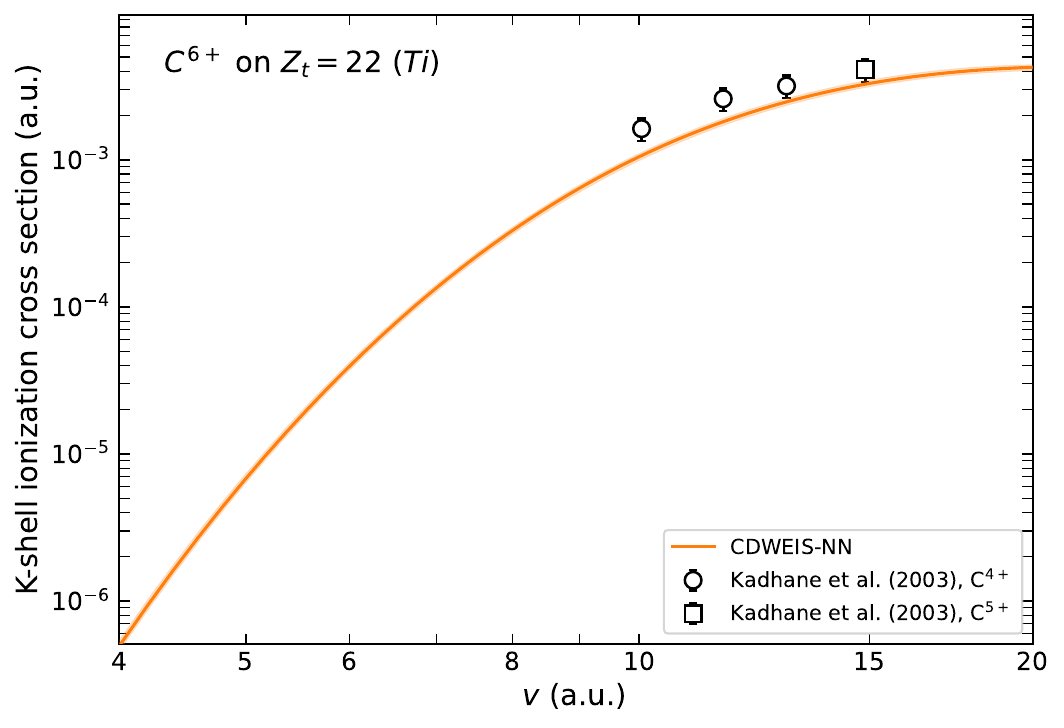}
\includegraphics[width=0.47\textwidth]{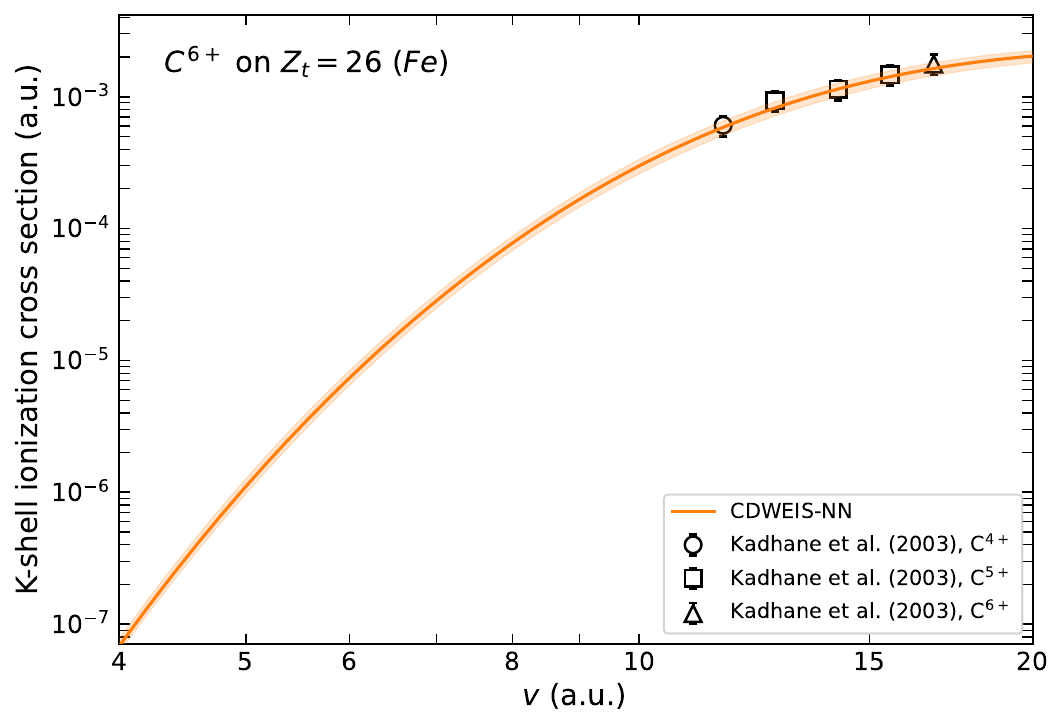}\\
\includegraphics[width=0.47\textwidth]{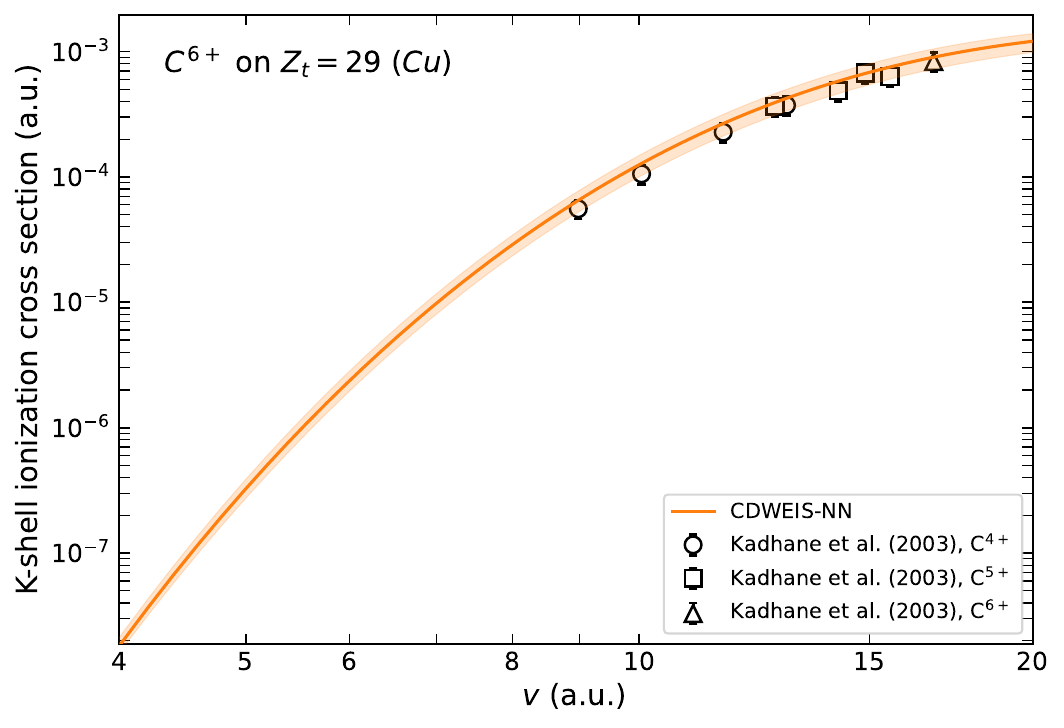}
\includegraphics[width=0.47\textwidth]{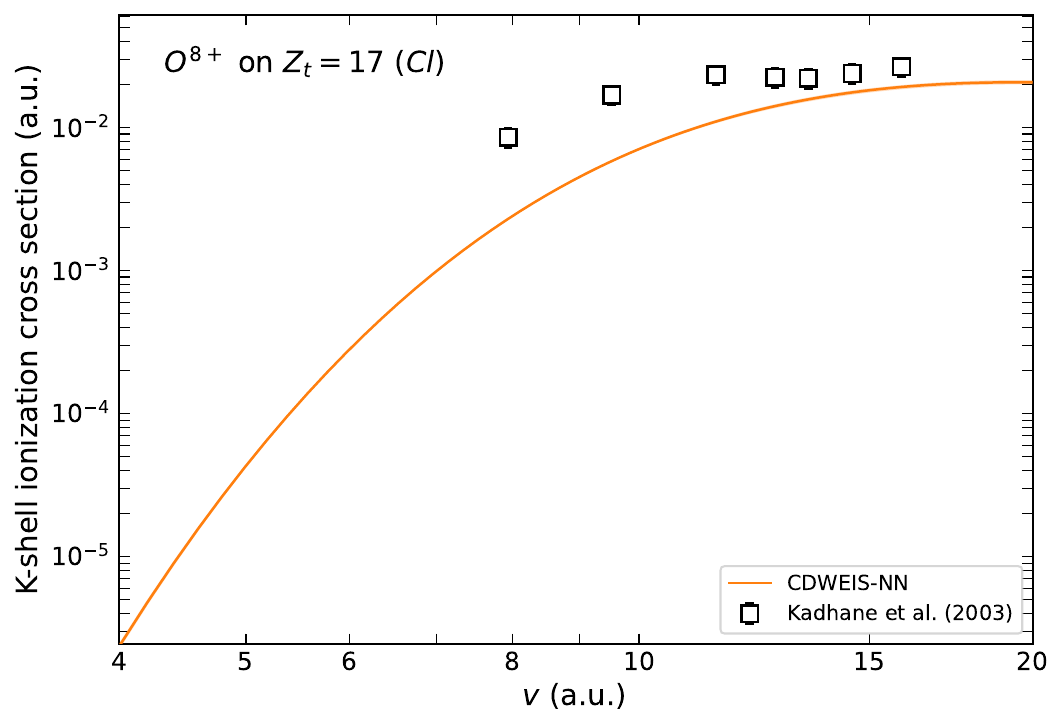}\\
\includegraphics[width=0.47\textwidth]{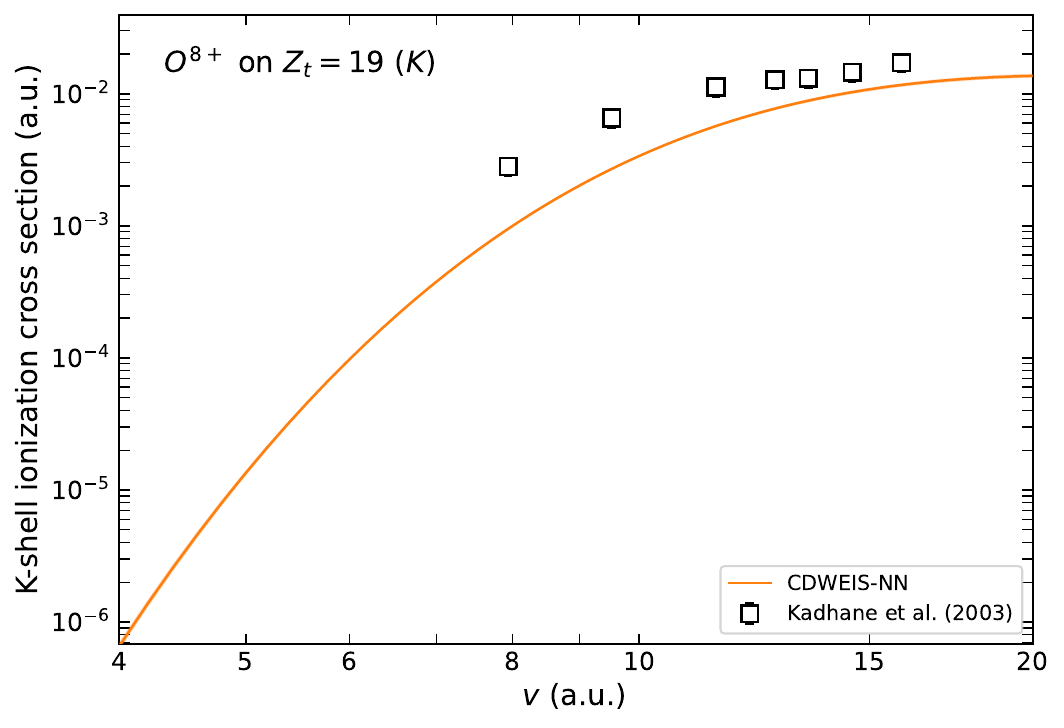}
\includegraphics[width=0.47\textwidth]{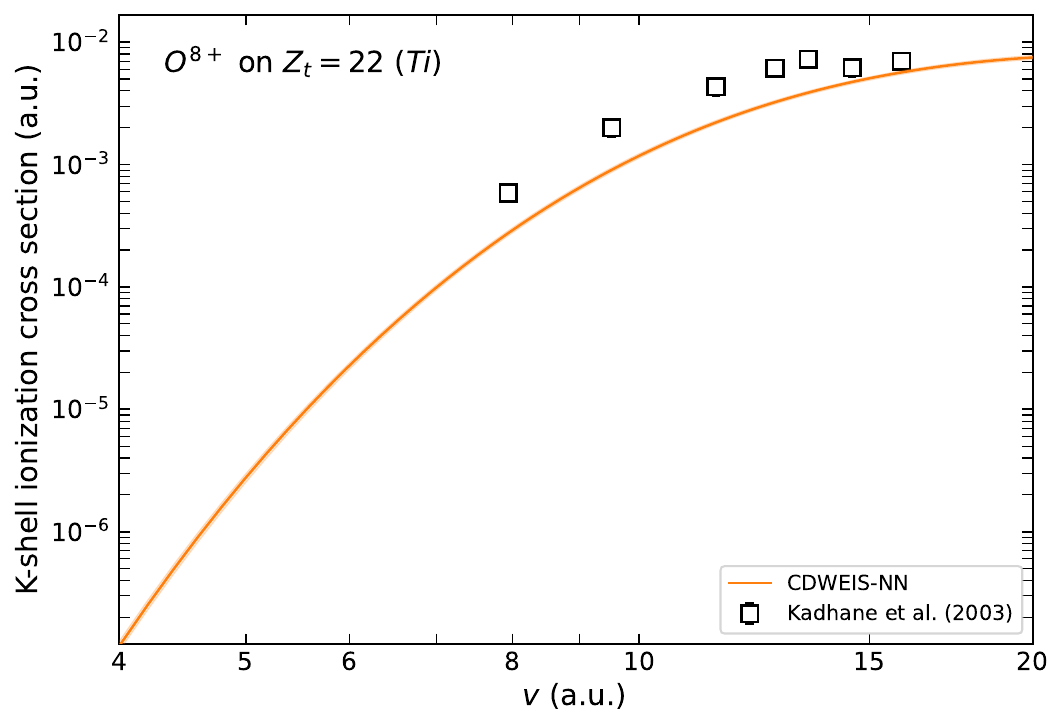}\\
\includegraphics[width=0.47\textwidth]{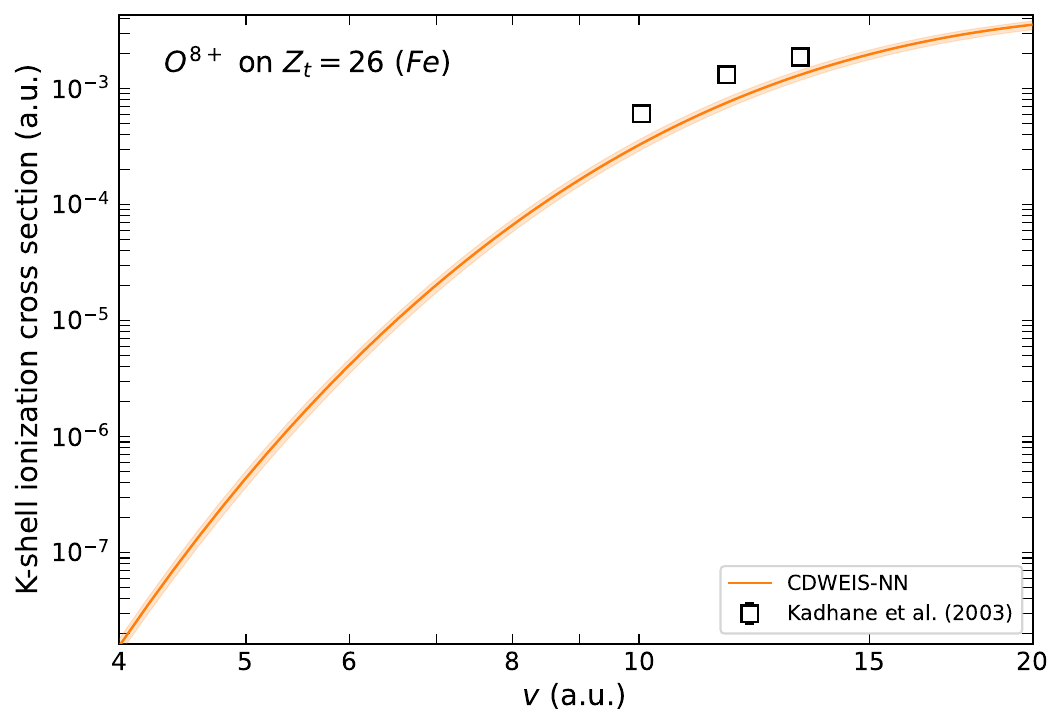}
\includegraphics[width=0.47\textwidth]{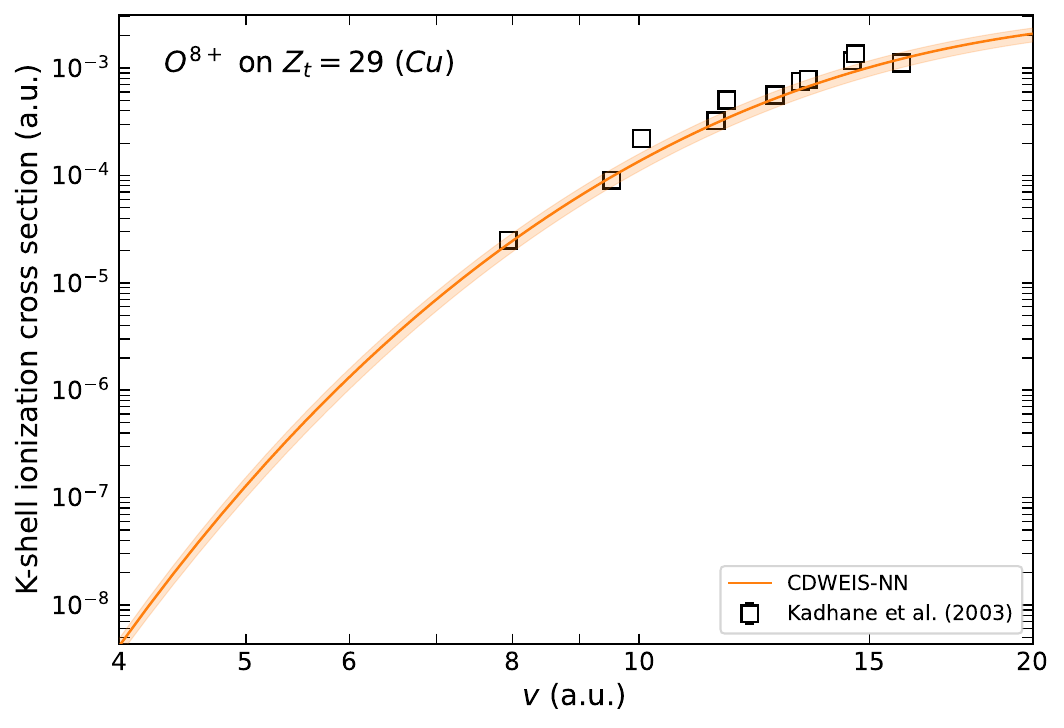}
\caption{K-shell ionization cross sections for C$^{6+}$ and O$^{8+}$ impact on Cl, K, Ti, Fe, and Cu. Lines show CDWEIS-NN predictions for projectile--target pairs absent from the CDW-EIS database. 
Shaded bands indicate the 16th--84th percentiles of the predictions of the five ensemble members and should not be interpreted as confidence intervals. Symbols show the experimental K-shell ionization cross sections derived from K X-ray measurements on solid targets: open circles, squares, and triangles identify the C$^{4+}$, C$^{5+}$, and C$^{6+}$ data of Ref.~\cite{KadhaneC6_2003}, and the oxygen data of Ref.~\cite{KadhaneO8_2003} correspond to projectiles with filled K shells. Error bars show typical experimental uncertainties ($17.5\%$ for carbon and $15\%$ for oxygen).}
\label{fig:C6O8Kshell}
\end{figure}

Figure~\ref{fig:C6O8Kshell} compares the CDWEIS-NN predictions for K-shell ionization of Cl, K, Ti, Fe, and Cu by C$^{6+}$ and O$^{8+}$ projectiles with the experimental cross sections derived from K X-ray measurements on solid targets by Kadhane \textit{et al.}~\cite{KadhaneC6_2003,KadhaneO8_2003}. 
The carbon data were obtained with C$^{4+}$, C$^{5+}$, and C$^{6+}$ ions, and the oxygen data with projectiles with no vacancy in the K shell. 
The experimental data indicate that the direct K-shell ionization cross sections are largely independent of the incident charge state, that is, of the number of electrons carried by the projectile.
K-shell ionization is dominated by close collisions, in which the projectile electrons barely screen the nucleus. The weak dependence on the incident charge state justifies comparing the measurements with predictions for bare projectiles. At the projectile velocities covered by these experiments, capture into higher projectile shells is negligible.
 
None of the corresponding projectile--target pairs was explicitly included in the CDW-EIS training database. 
The figure therefore probes the practical use of the neural-network emulator for obtaining cross sections within the covered projectile-charge and target-atomic-number domain, but for previously uncalculated collision systems.

The predicted curves reproduce the overall magnitude and energy dependence of the measured K-shell cross sections across the five targets and for both projectile species. This comparison cannot isolate the neural-network approximation from the accuracy of the underlying CDW-EIS description, since direct CDW-EIS calculations are not available for these pairs. It nevertheless provides the most direct external test of the complete predictive framework: a CDW-EIS-informed neural-network prediction is consistent with measurements for collision systems absent from the explicit theoretical database.

%%%%%%%%%%%%%%%%%%%%%%%%%%%%%%%%%%%%%%%%%%%%%%%%%%%%%%%%%%%%%%%%%%%%%%%%
%%%%%%%%%%%%%%%%%%%%%%%%%%%%%%%%%%%%%%%%%%%%%%%%%%%%%%%%%%%%%%%%%%%%%%%%
\section{Conclusions}
\label{sec:conclusions}

We have presented a CDW-EIS database of subshell-resolved total ionization cross sections for 132 projectile--target systems. The database covers antiproton impact and bare-ion impact over a broad range of projectile charges, for atomic targets extending from H to Kr. It contains 6239 tabulated values for all occupied $nl$ subshells over a broad range of projectile velocities and is openly available through the online repository.

Comparison with available measurements for total ionization of H and He and for K-  and L-shell ionization confirms the reliability of the direct CDW-EIS calculations. Across the selected benchmarks, the calculations reproduce the absolute scale and the velocity dependence of the measured cross sections, including the low-velocity charge-sign effects in proton and antiproton collisions. This agreement verifies the present database calculations; the CDW-EIS model itself has been extensively tested in previous studies.

To extend the practical coverage of the direct calculations, we constructed the five-member CDWEIS-NN ensemble. Its group-held-out validation, in which complete projectile--target pairs are withheld from individual members, gives a median relative error of $3.1\%$ and places $99.5\%$ of the held-out cross sections within $30\%$ of the CDW-EIS values. The final ensemble also reproduces the direct calculations and gives satisfactory agreement with the experimental benchmarks considered in this work. The ensemble spread is useful as an indicator of the consistency among individual network predictions, but it should not be interpreted as a complete uncertainty estimate for the underlying collision physics.

The database and its neural-network surrogate thus provide complementary resources: the former supplies the underlying \textit{ab initio} cross sections and their physical systematics, while the latter enables fast prediction for systems that have not yet been calculated directly. Future extensions will incorporate additional projectile--target combinations and collision energies, and the network will be retrained accordingly. A different goal, beyond the scope of the present work, is a network describing the ionization cross sections themselves rather than a specific collision model, which could combine several theoretical approaches and experimental data and incorporate physical constraints into the learning procedure.

%%%%%%%%%%%%%%%%%%%%%%%%%%%%%%%%%%%%%%%%%%

%%%%%%%%%%%%%%%%%%%%%%%%%%%%%%%%%%%%%%%%%%
\vspace{6pt} 

%%%%%%%%%%%%%%%%%%%%%%%%%%%%%%%%%%%%%%%%%%%%%%%%%%%%%%%%%%%%%%%%%%%%%%%%
%%%%%%%%%%%%%%%%%%%%%%%%%%%%%%%%%%%%%%%%%%%%%%%%%%%%%%%%%%%%%%%%%%%%%%%%
\section*{Author Contributions}
All authors contributed equally to the development of this work

%%%%%%%%%%%%%%%%%%%%%%%%%%%%%%%%%%%%%%%%%%%%%%%%%%%%%%%%%%%%%%%%%%%%%%%%
%%%%%%%%%%%%%%%%%%%%%%%%%%%%%%%%%%%%%%%%%%%%%%%%%%%%%%%%%%%%%%%%%%%%%%%%
\section*{Funding}
We acknowledge the PIP 11220250100903CO grant from CONICET. 

%%%%%%%%%%%%%%%%%%%%%%%%%%%%%%%%%%%%%%%%%%%%%%%%%%%%%%%%%%%%%%%%%%%%%%%%
%%%%%%%%%%%%%%%%%%%%%%%%%%%%%%%%%%%%%%%%%%%%%%%%%%%%%%%%%%%%%%%%%%%%%%%%
\section*{Acknowledgments}
In a context of severe constraints for the Argentine universities and scientific system, the persistence of the scientific community has been essential to keep this kind of research alive. 
We thank our colleagues for that collective effort.

%%%%%%%%%%%%%%%%%%%%%%%%%%%%%%%%%%%%%%%%%%%%%%%%%%%%%%%%%%%%%%%%%%%%%%%%
%%%%%%%%%%%%%%%%%%%%%%%%%%%%%%%%%%%%%%%%%%%%%%%%%%%%%%%%%%%%%%%%%%%%%%%%
\section*{Data Availability}
The complete
\href{https://www.iafe.uba.ar/u/dario/cdwbase/}{CDW-EIS database} is publicly
available. The trained CDWEIS-NN ensemble is publicly available through
\href{https://colab.research.google.com/drive/1TrblWWwpTEwSlGh--dSyBWkIYNvUbksg}{\texttt{CDWEISNN.ipynb}}, a self-contained Google Colab
notebook. No local software installation is required: users can specify a
projectile--target pair $(Z_\mathrm{P}, Z_\mathrm{T})$ and obtain the
corresponding subshell-resolved and total ionization cross sections, tables,
and figures directly in a web browser.

%%%%%%%%%%%%%%%%%%%%%%%%%%%%%%%%%%%%%%%%%%%%%%%%%%%%%%%%%%%%%%%%%%%%%%%%
%%%%%%%%%%%%%%%%%%%%%%%%%%%%%%%%%%%%%%%%%%%%%%%%%%%%%%%%%%%%%%%%%%%%%%%%
\section*{Conflicts of Interest}
The authors declare no conflicts of interest.  

%%%%%%%%%%%%%%%%%%%%%%%%%%%%%%%%%%%%%%%%%%%%%%%%%%%%%%%%%%%%%%%%%%%%%%%%
%%%%%%%%%%%%%%%%%%%%%%%%%%%%%%%%%%%%%%%%%%%%%%%%%%%%%%%%%%%%%%%%%%%%%%%%
%\reftitle{References}

%=====================================
% References, variant A: external bibliography
%=====================================

\bibliographystyle{unsrt}
\bibliography{CDWEISreferences}

%%%%%%%%%%%%%%%%%%%%%%%%%%%%%%%%%%%%%%%%%%%%%%%%%%%%%%%%%%%%%%%%%%%%%%%%
%%%%%%%%%%%%%%%%%%%%%%%%%%%%%%%%%%%%%%%%%%%%%%%%%%%%%%%%%%%%%%%%%%%%%%%%
\end{document}